\documentclass{aa}  

\usepackage{hyperref}

\usepackage{amsmath}
\usepackage{graphicx}
\usepackage{graphics}
\usepackage{subcaption}
\usepackage{natbib}{}
\usepackage{xcolor}
\usepackage{url}
\usepackage{hyperref}

\bibpunct{(}{)}{;}{a}{}{,}

\def\teff{\ifmmode T_{\rm eff} \else $T_{\mathrm{eff}}$\fi}

\def\ltsima{$\buildrel<\over\sim$}
\def\lsim{\lower.5ex\hbox{\ltsima}}

\newcommand{\hii}{H~{\sc ii}}
\newcommand{\ha}{\ifmmode {\rm H}\alpha \else H$\alpha$\fi}
\newcommand{\hb}{\ifmmode {\rm H}\beta \else H$\beta$\fi}
\newcommand{\hg}{\ifmmode {\rm H}\gamma \else H$\gamma$\fi}
\newcommand{\lya}{\ifmmode {\rm Ly}\alpha \else Ly$\alpha$\fi}

\newcommand{\Heiiuv}{He~{\sc ii} $\lambda$1640}

\newcommand{\ebv}{\ifmmode E_{\rm B-V} \else $E_{\rm B-V}$\fi}
\newcommand{\av}{\ifmmode A_{\rm V} \else $A_{\rm V}$\fi}

\def\micron{$\mu$m}

\def\cmc{cm$^{-3}$}

\def\msun{\ifmmode M_{\odot} \else M$_{\odot}$\fi}
\def\msunyr{\ifmmode M_{\odot} {\rm yr}^{-1} \else M$_{\odot}$ yr$^{-1}$\fi}
\def\zsun{\ifmmode Z_{\odot} \else Z$_{\odot}$\fi}

\def\lsun{\ifmmode L_{\odot} \else L$_{\odot}$\fi}

\def\mup{\ifmmode M_{\rm up} \else M$_{\rm up}$\fi}
\def\mlow{\ifmmode M_{\rm low} \else M$_{\rm low}$\fi}
\def\aap{A\&A}

\def\apjl{ApJL}
\def\apjs{ApJS}

\newcommand{\oh}{\ifmmode 12 + \log({\rm O/H}) \else$12 + \log({\rm
O/H})$\fi}

\newcommand{\oii}{[O~{\sc ii}]}
\newcommand{\oiii}{[O~{\sc iii}]}

\def\Oii{[O~{\sc ii}] $\lambda$3727}
\def\Oiii{[O~{\sc iii}] $\lambda\lambda$4959,5007}
\def\Oiiib{[O~{\sc iii}]$\lambda 5007$}
\def\Oiiit{[O~{\sc iii}]$\lambda 4363$}
\newcommand{\Neiii}{[Ne~{\sc iii}] $\lambda$3869}

\def\flyf{\ifmmode f_{\rm Lyf} \else $f_{\rm Lyf}$\fi}
\def\pz{\ifmmode P(z) \else $P(z)$\fi}
\def\ki2{\ifmmode \chi^2 \else $\chi^2$\fi}
\def\zphot{\ifmmode z_{\rm phot} \else $z_{\rm phot}$\fi}

\newcommand{\xphot}{\ifmmode x_\gamma \else $v_\gamma$\fi}
\newcommand{\xobs}{\ifmmode x_{\rm obs} \else $x_{\rm obs}$\fi}
\newcommand{\xcmf}{\ifmmode x_{\rm CMF} \else $x_{\rm CMF}$\fi}
\newcommand{\vexp}{\ifmmode V_{\rm exp} \else $V_{\rm exp}$\fi}
\newcommand{\vmax}{\ifmmode V_{\rm max} \else $V_{\rm max}$\fi}
\newcommand{\nh}{\ifmmode N_{\rm HI} \else $N_{\rm HI}$\fi}
\newcommand{\dv}{\ifmmode \Delta v({\rm em-abs}) \else $\Delta v({\rm em}-{\rm abs})$\fi}

\def\fesc{\ifmmode f_{\rm esc} \else $f_{\rm esc}$\fi}
\def\fescrel{\ifmmode f_{\rm esc,rel} \else $f_{\rm esc,rel}$\fi}

\def\frellya{\ifmmode f^{\rm rel}_{\rm{Ly}\alpha} \else $f^{\rm rel}_{\rm{Ly}\alpha}$\fi}

\def\hii{H{\sc ii}}

\newcommand{\mstar}{\ifmmode M_\star \else $M_\star$\fi}
\newcommand{\muv}{\ifmmode M_{\rm UV} \else $M_{\rm UV}$\fi}
\newcommand{\auv}{\ifmmode A_{\rm UV} \else $A_{\rm UV}$\fi}
\newcommand{\luv}{\ifmmode L_{\rm UV} \else $L_{\rm UV}$\fi}
\newcommand{\lir}{\ifmmode L_{\rm IR} \else $L_{\rm IR}$\fi}
\newcommand{\lbol}{\ifmmode L_{\rm bol} \else $L_{\rm bol}$\fi}
\newcommand{\liruv}{\ifmmode L_{\rm IR+UV} \else $L_{\rm IR+UV}$\fi}
\newcommand{\liroveruv}{\ifmmode L_{\rm IR}/L_{\rm UV} \else $L_{\rm IR}/L_{\rm UV}$\fi}
\newcommand{\nlyc}{\ifmmode N_{\rm Lyc} \else $N_{\rm Lyc} $\fi}
\newcommand{\rholyc}{\ifmmode \rho_{\rm Lyc} \else $\rho_{\rm Lyc} $\fi}
\newcommand{\chion}{\ifmmode \xi_{\rm ion} \else $\xi_{\rm ion}$\fi}
\newcommand{\chioncorr}{\ifmmode \xi_{\rm ion}^0 \else $\xi_{\rm ion}^0$\fi}

\newcommand{\Civuv}{C~{\sc iv} $\lambda$1550}
\newcommand{\Civ}{C~{\sc iv}}
\newcommand{\Ciii}{C~{\sc iii}]}
\newcommand{\Ciiiuv}{C~{\sc iii}] $\lambda$1909}
\newcommand{\Oiiiuv}{O~{\sc iii}] $\lambda$1666}

\newcommand{\Niiiuv}{N~{\sc iii}] $\lambda$1750}
\newcommand{\Nivuv}{N~{\sc iv}] $\lambda$1486}
\newcommand{\Niii}{N~{\sc iii}]}
\newcommand{\Niv}{N~{\sc iv}]}

\begin{document}

\title{Discovery of new N-emitters over a wide redshift range}
\subtitle{}
\author{I. Morel\inst{1}\thanks{E-mail: Ilona.Morel@unige.ch},  
D. Schaerer\inst{1,2}, 
R. Marques-Chaves\inst{1},
N. Prantzos\inst{3},
C. Charbonnel\inst{1,2},
G. Brammer\inst{4},
M. Xiao\inst{1},
M. Dessauges-Zavadsky\inst{1}
}
  \institute{Observatoire de Gen\`eve, Universit\'e de Gen\`eve, Chemin Pegasi 51, 1290 Versoix, Switzerland
\and CNRS, IRAP, 14 Avenue E. Belin, 31400 Toulouse, France
\and Institut d'Astrophysique de Paris, UMR 7095 CNRS, Sorbonne Université, 98bis, Bd Arago, 75014 Paris, France
\and Cosmic Dawn Center (DAWN), Niels Bohr Institute, University of Copenhagen, Jagtvej 128, K\o benhavn N, DK-2200, Denmark
}

\authorrunning{Morel et al.}
\titlerunning{Discovery of new N-emitters}

\date{Received date; accepted date}

\abstract{Recent James Webb Space Telescope (JWST) observations have revealed a small number of galaxies with UV spectra exhibiting intense emission lines of nitrogen, which are not typically detected in galaxy spectra. The observations indicate super-solar N/O abundances at low metallicity, whose nature is currently intensly debated. }
{To better understand these enigmatic objects and provide new constraints on proposed scenarios, we have undertaken a systematic search for galaxies with UV emission lines of nitrogen (\Nivuv\ and/or \Niiiuv). }
{Using public JWST NIRSpec data, we have identified 45 N-emitters with robust \Niii\ or \Niv\ detections, including 4 previously known objects. We then classified these objects, determined the relative chemical abundance ratios of C, N, O, and H, determined other properties, and carried out a statistical analysis of the N-emitter population.  }
{We find N-emitters from redshift $z \sim 3 - 11$ among a broad diversity of galaxies, in terms of morphology, UV magnitude, stellar mass, SFR, metallicity, and rest-optical emission line strengths.
The UV nitrogen lines show typical equivalent widths between $\sim 5$ \AA\ and up to $\sim 100$ \AA\ in few cases. Diverse ionisation conditions, as traced by \Nivuv/\Niiiuv, are observed. Carbon lines (\Civuv\ and \Ciiiuv) are generally fainter than the N lines. 
Using strong line calibrations established at high-redshift, we find metallicities $\oh \sim 7.15-8.5$, including thus also high metallicities.
The \hb\ equivalent width of N-emitters varies strongly,
and sources with low EWs show clear signs of a Balmer break, indicative of composite stellar populations combining both young ($\protect\la 10$ Myr) stars responsible of the UV emission lines and an older population contributing to the rest-optical spectrum. 
Supersolar N/O ratios are found in all N-emitters. C/O abundances are comparable to those of galaxies at the same metallicity, and all N-emitters show high N/C ratios or lower limits ($\log($N/C$) \protect\ga 0.5$), independently of metallicity. The observed abundance ratios are compatible with ejecta from H-burning and do not show signs of Carbon enhancements, even at higher metallicities. 
Finally, we find that the fraction of N-emitters increases with redshift, and we quantify this evolution. }
{Our study increases the sample of known N-emitters by a factor $\sim 3$, reveals a diversity of properties among N-emitters, and provides new constraints on their nature.}

\keywords{Galaxies: abundances -- Galaxies: high-redshift -- Galaxies: ISM}

\maketitle

\section{Introduction}
\label{s_intro}

JWST observations have revealed an unexpected diversity of rest-frame UV emission lines in galaxies at the earliest cosmic epochs. More specifically, GN-z11 at $z=10.6$ exhibits a very peculiar NIRSpec spectrum \cite{Bunker2023JADES-NIRSpec-S} characterised by strong \Nivuv\ and \Niiiuv\ emission lines, which are very rare in typical star-forming galaxies. Using emission line measurements, \cite{2023Cameron} derived an exceptionally high nitrogen-to-oxygen ratio for GN-z11 ($\log($N/O$)\ga\ -0.25$) for such low metallicity ($\oh \approx 7.90$, \citealt{2025Alvarez}). 
This finding is at odds with standard galactic chemical evolution models and observations of low-redshift galaxies and \hii\ regions, where the N abundance is essentially constant at low metallicities and increases only later, when the bulk of nitrogen -- produced as a secondary element by intermediate-mass AGB stars -- comes into play \citep[][]{Henry2000On-the-Cosmic-O,Kobayashi2025Nucleosynthesis}. 
Following this discovery, a few other objects that deviate significantly from the standard N/O versus O/H relation have been reported
\citep{2024MarquesChaves,2024Schaerer,2025Naidu,2024Navarro,2024Castellano}.
These and other nitrogen-enhanced galaxies now include $\sim22$ objects, as compiled by \citet{Ji2025Connecting-JWST}, predominantly found at high redshift, and several metal-poor low-$z$ galaxies with enhanced N/O abundances have also been reported recently \citep[see][]{Bhattacharya2025The-origin-of-e,Martinez2025Under-Pressure:}.

The newly discovered high-$z$ objects share extreme physical properties beyond their chemical "anomalies". Their C/O ratio remains surprisingly consistent with those of "normal" star-forming galaxies \cite{2024MarquesChaves,Ji2025Connecting-JWST}. They typically exhibit high electron densities \citep{2024bJi}, strong UV luminosities, and remarkably elevated surface densities of both star formation rate (SFR) and stellar mass (\mstar) \citep{2024Schaerer}. Furthermore, these galaxies show relatively high $\mathrm{EW(H\beta)}$ \citep{Topping2025Deep-Rest-UV-JW}, suggesting the presence of young stellar populations. 
These shared properties depict a compact, dense and young star-forming environment within these peculiar objects.
And the detection of high N/O ratios in high-redshift galaxies suggests that the responsible process must occur relatively rapidly. 

A variety of scenarios have been proposed to explain the observed N-enhancement, including enrichment from massive star winds (WR stars), AGB stars and selective outflows, or possibly more ``exotic'' scenarios such as tidal disruption of stars from encounters with black holes, very massive stars (VMS) ejecta or even supermassive stars (SMS) \citep{Bunker2023JADES-NIRSpec-S,2023Cameron,Charbonnel2023N-enhancement-i, DAntona2023,Bhattacharya2025The-origin-of-e,Vink2023A&A...VMS, 2024MarquesChaves, DAntona2025, McClymont2025The-THESAN-ZOOM}.
However, the origin of the observed N-enrichment remains largely unanswered. 

To better understand these interesting objects, larger samples and new observational constraints are needed. To do so, we have used a large collection of public JWST spectra and carried out a systematic search for N-emitters, which we define as galaxies showing UV emission lines of \Nivuv\ and/or \Niiiuv.
Our study significantly increases the number of N-emitters at $z \sim 3-10$, revealing a diversity of galaxies among them, and providing new insights on these enigmatic objects. 

The paper is structured as follows. In Sect.\ref{s_obs}, we describe the observational data and the new N-emitter sample. Sect.\ref{s_sample} describes empirical and basic derived properties of the sample.
We present the derived O/H metallicities and chemical abundances, as well as their morphological properties in Sect.\ref{s_props}. Finally, we discuss the surprising diversity of the N-emitter population and provide a statistical estimate of their incidence, before discussing the potential origins of the observed nitrogen enrichment (Sect.\ref{s_discuss}). Our main results are summarised in Sect.~\ref{s_conclude}. Throughout this work, we assume a concordance cosmology with $\mathrm{\Omega_{m}=0.3}$, $\mathrm{\Omega_{\Lambda}=0.7}$ and $\mathrm{H_{0}=70 \ km\ s^{-1} Mpc^{-1}}$.


\section{JWST observations and sample selection}
\label{s_obs}

To search for N-emitter candidates, we have used the DAWN JWST Archive (DJA)\footnote{\url{https://dawn-cph.github.io/dja}} database for distant galaxies and its version, which includes automated measurements of numerous emission lines (\citealt{msaexp}, \citealt{Heintz2025The-JWST-PRIMAL}, \citealt{Valentino2025}, and Brammer, Private Communication). 
We have selected galaxies with reliable spectroscopic redshifts ({\it grade=3}) and we extracted all the spectra where one or both of the UV lines of \Nivuv\ and \Niiiuv\ are potentially detected, with a signal-to-noise ratio $S/N \geq 3$. 
To have the broadest possible spectral range, we use the NIRSpec PRISM spectra only, which cover wavelengths $\lambda_{\rm obs} \sim 0.7 - 5.2$ \micron, with a resolution varying from $R \simeq 30$ to $\simeq 300$ over this range.
Our selection covers all MSA PRISM spectra included in the DJA database until August 18, 2025. 
The preselection yielded 633 automatically-selected candidates.

\subsection{Emission line measurements and final selection}

After assessing the quality of the N {\sc iv}] and N {\sc iii}] detections, we found that several of them are spurious. 
To obtain a more reliable estimate of the detection significance of the UV nitrogen emission lines, we remeasured their fluxes, equivalent widths (EWs), and line widths (FWHM) using two different approaches, described in the Appendix \ref{s_app_measure}.  

We define our sample of N-emitters as sources with $S/N \geq 3$ in the N {\sc iv}] and/or N {\sc iii}] lines, identified using both \texttt{Lime} and our custom-made code. This yields a total of 70 N-emitter candidates. After detailed visual inspection of the 1D and 2D spectra by several team members, we excluded 24 additional sources from the sample, as nitrogen lines show significant deviations from the expected line positions, issues in the 2D spectra, or features consistent with spurious detections. Our final selection consists of 45 N-emitters, listed in Table~\ref{tab_nemm}. 

\begin{figure}[tb]
\includegraphics[width=0.5\textwidth]{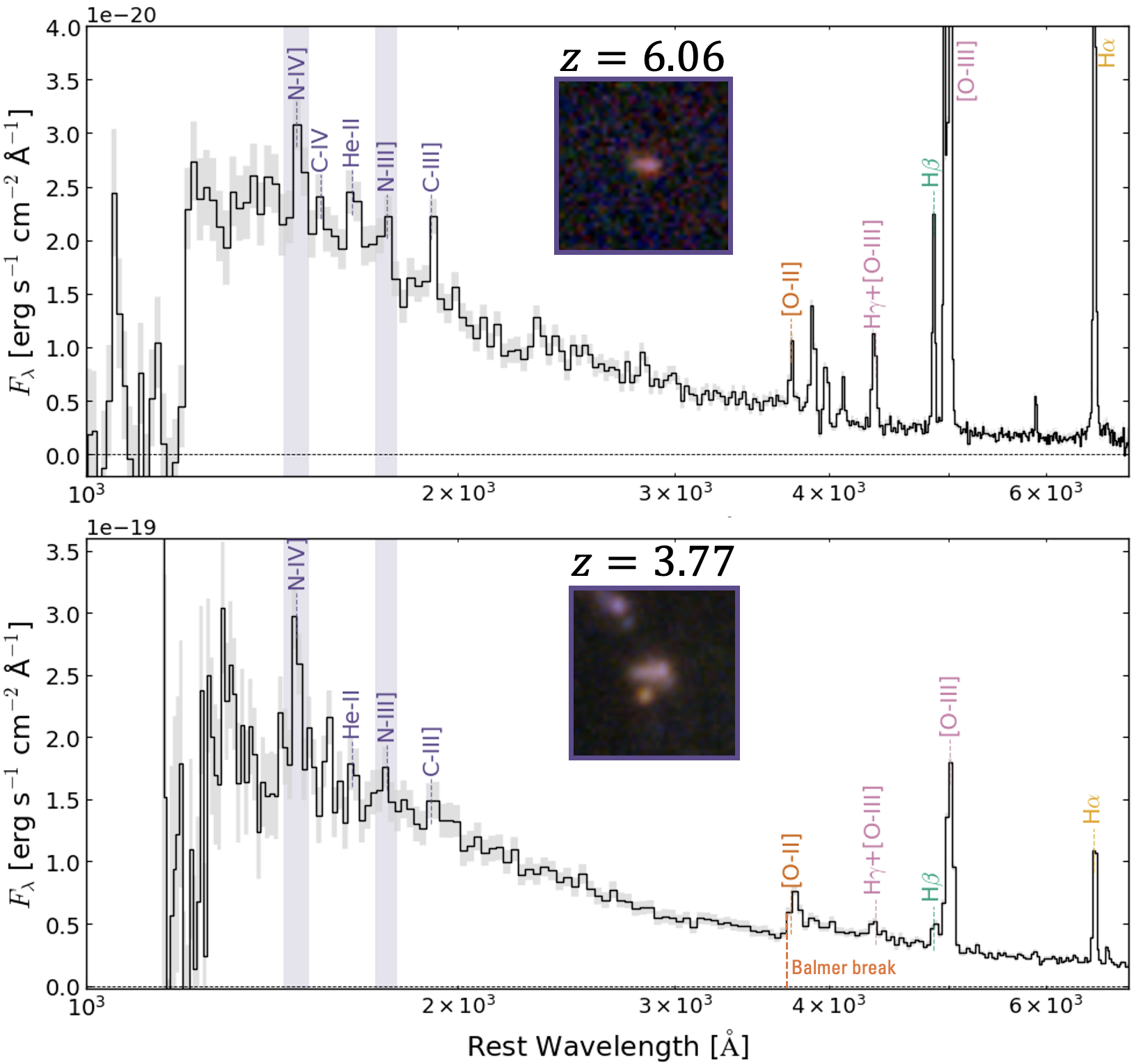}
\caption{NIRSpec/PRISM spectra of two N-emitters, illustrating the diversity of our sample. The top spectrum corresponds to a highly star-forming galaxy with strong optical lines. The bottom one shows a Balmer break at $\sim 3740$\AA, a very low EW(H$\beta$), and faint optical lines. Both show at least one significant UV nitrogen line.}
\label{fig_spec}
\end{figure}

\subsection{Spectral classification: SFG, AGN, and LRD}

Our sample of 45 new N-emitters exhibits a wide range of spectral features, as highlighted in Figure \ref{fig_spec} and discussed below (Sect.~\ref{s_sample}). 
We inspected the spectra of all sources to identify potential AGN through the presence of broad emission profiles in Balmer lines or the presence of high-ionisation features (e.g., [Ne\,{\sc v}]~$\lambda$3346) and for UV and optical line ratios indicative of AGN activity. 
Our procedure and the resulting classification (star-forming (SF), AGN, or Little Red Dot) are described in Appendix~\ref{s_app_agn}. In short, for the majority of sources (36 out of 45) we do not find any signs of AGN or LRD.

\begin{figure*}[tb]
\includegraphics[width=0.5\textwidth]{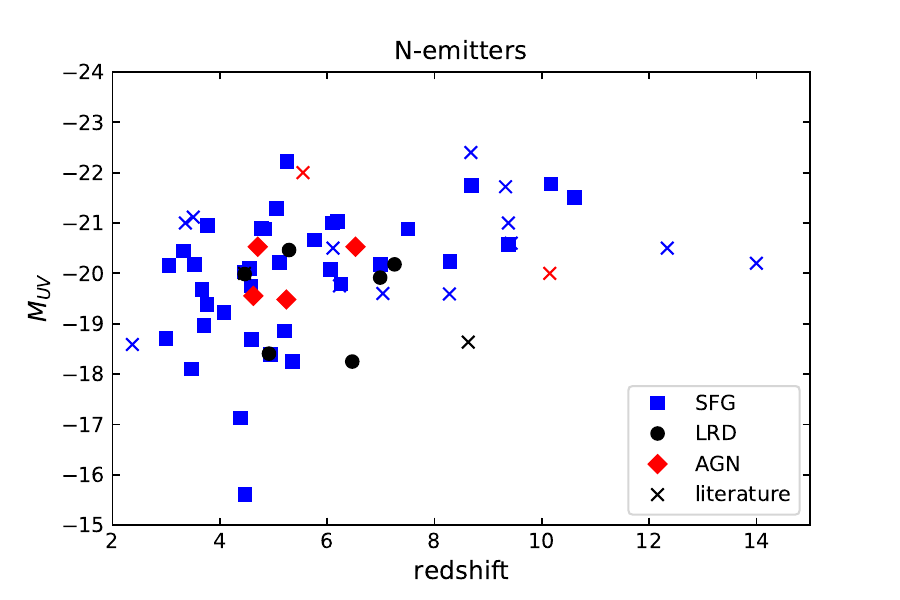}
\includegraphics[width=0.5\textwidth]{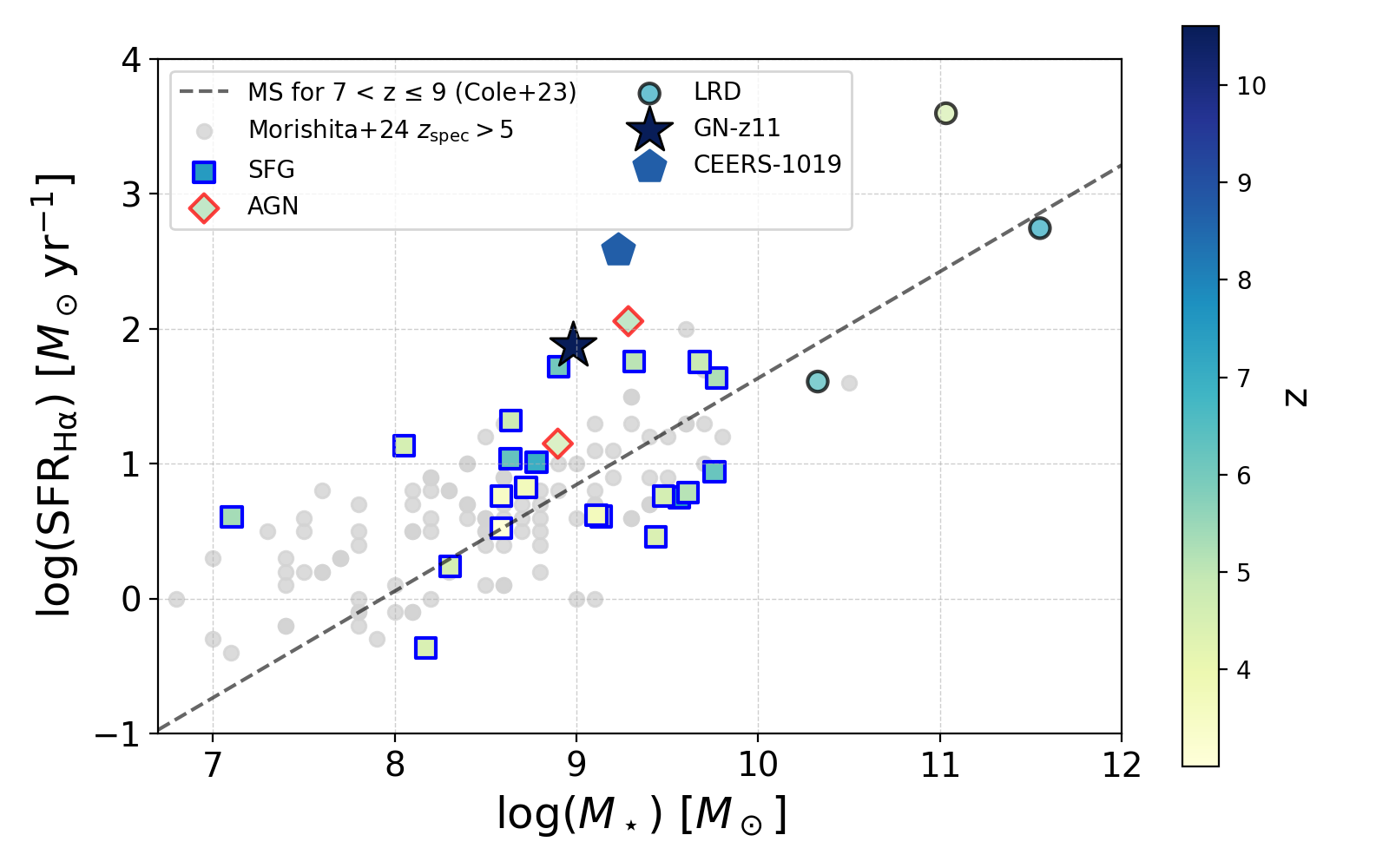}
\caption{{\em Top:} Absolute UV magnitude \muv\ of all known N-emitters (60 in total) versus redshift.
The literature sample is marked with crosses. Colours distinguish objects categorised as SFGs (blue), LRDs (black), and AGN (red). The literature sample also includes four lensed galaxies, whose magnitudes have been corrected for gravitational lensing.
{\em Bottom:} Mass--SFR showing a subset of the N-emitters (symbols colour-coded by redshift) for which the stellar mass could be derived, and the SF main sequence of \citet{2025Cole} at $z=7-9$ (dashed).}
\label{fig_muv}
\end{figure*}

\subsection{Extinction correction}\label{s_extinction}

We measure the colour excess, \ebv, from the Balmer decrement using the main Balmer lines, from H$\alpha$ to H$\delta$, and adopting the extinction curve from \citet{Cardelli1989The-relationshi} assuming standard nebular conditions ($n_e = 250$\,cm$^{-3}$ and $T_e = 10^{4}$\,K). Owing to the limited PRISM resolution, the H$\alpha$ and H$\gamma$ lines are blended with emission from [N\,{\sc ii}] $\lambda\lambda$6550,6585 and [O\,{\sc iii}] $\lambda$4363, respectively. We therefore assume that the contribution from these neighbouring lines amounts to roughly $\simeq 10\%$ of the observed fluxes. 

We calculate all available Balmer line ratios and compare them to the predicted values for a given \ebv\ and the adopted extinction curve. Our fiducial \ebv\ values are those that minimize $\chi^{2}$. 
For 20 sources, we find non-physical Balmer decrements, i.e., consistent with negative \ebv, although the uncertainties for several of them are consistent with positive reddening. For these cases, we adopt \ebv$=0$. Sources with physical \ebv\ values exhibit a wide range of colour excesses, from $\simeq0.03$\,mag up to $\simeq2.7$\,mag, with LRDs being the most obscured. Finally, the spectra of nine sources show no or only one significantly detected Balmer line, and we conservatively assume \ebv$=0$ for these objects. 
The full sample has a mean extinction of $E(B-V) = 0.216^{+0.285}_{-0.146}$ and a median value close to zero. Finally, the rest-UV and optical emission line fluxes are corrected for extinction, adopting the \cite{Reddy2016Spectroscopic-M} curve.

\subsection{Literature samples}

Our selection recovers several (4) previously known N-emitters reported in the literature:
GN-z11, CEERS-1019, GN-z9p4, and UNCOVER-45924 (see \citealt{Bunker2023JADES-NIRSpec-S}; \citealt{2024MarquesChaves}; \citealt{2024Schaerer} and \citealt{2024Labbe}). 
%
The recent compilation of \cite{Ji2025Connecting-JWST} includes 13 objects showing emission in \Nivuv\ or \Niiiuv, which we define here as N-emitters\footnote{They also include other objects with high N/O ratios, for which currently no rest-UV spectra are available.}
Three N-emitters were reported after that \citep[MoM-z14, UNCOVER-3686, CANUCS-LRD-z8.6 from][]{2025Naidu, Fujimoto2024ApJ...977..250F_UNCOVER, Tang2025The-JWST-Spectr, Morishita2025A-Nitrogen-rich}, and two other lensed N-emitters at $z \sim 3.4-3.5$, SMACS 2031 and the Lynx arc, were discovered earlier \citep[see][]{2024MarquesChaves,Patricio2016A-young-star-fo,Fosb03}.
Among those 18 N-emitters, 11 have MSA/PRISM observations, and 4 have been observed with medium or high-resolution gratings. Our selection has not recovered 7 objects with MSA/PRISM, which are GN-z8-LAE, GS-z9-0, GHZ9, GHZ2=GLASS-z12, MoM-z14, UNCOVER-3686, CANUCS-LRD-z8.6 reported by \citet{2024Navarro}, \citet{2025Curti}, \citet{2024Napolitano}, \citet{2024Castellano}, \citet{2025Naidu}, \citet{Fujimoto2024ApJ...977..250F_UNCOVER}, and \citet{2024Tripodi}.
In several of them, the detection of N lines is tentative, and our procedure/measurements do not yield significant ($>3 \sigma$) and robust detections for these objects. We therefore do not add them to our sample, but subsequently compare our sample with this literature sample.

In short, combining the 18 previously known N-emitters and our sample (42 new objects), our search has more than tripled the number of these objects (to 60 N-emitters currently known).
Note also that \cite{Martinez2025Under-Pressure:} report 8 new objects at $z \la 0.04$ with N lines in the UV.

\section{A new sample of N-emitters and candidates}
\label{s_sample}


\subsection{Redshift, $M_{\rm UV}$, SFR, and stellar mass}
\label{s_sfr}

The distribution of redshift and UV brightness of the N-emitters is shown in the left panel of Fig.~\ref{fig_muv}.
The objects show a broad redshift distribution, from $z=14$ to $z \sim 3$, where the lower limit is determined by the minimum wavelength covered by the PRISM spectra ($\lambda_{\rm min} \sim 0.6$ \micron). 
Compared to the literature sample, our study adds many new objects at $z \sim 3-7$. It also shows that N-emitters are found among fainter galaxies than previously known, at least down to $\muv \sim -16$.

We estimated stellar masses and star-formation rates (SFRs) of a subset of N-emitters with available JWST NIRCam photometry and NIRSpec prism spectra using SED fitting with Bagpipes \citep{bagpipes}, with fixed spectroscopic redshifts. 
An up-to-second-order polynomial calibration is applied to correct residual slit-loss of the spectra compared to the photometry. We adopt a non-parametric star-formation history, BPASS stellar population models, the \citet{2018Salim} dust attenuation curve, and vary metallicity. Further methodological details will be presented elsewhere (Xiao et al., in prep.) 
We also derived the SFR from the dust-corrected H$\alpha$ (or \hb) line flux, following the relation from \citet{1998Kennicutt}.
The resulting stellar mass--SFR relation is shown in the right panel of Fig.~\ref{fig_muv}. Our sources broadly follow the main sequence (MS) of galaxies derived by \citet{2025Cole} in the redshift range $7 < z < 9$. This choice of redshift range is somewhat arbitrary, as the MS does not significantly change with redshift. 
Interestingly, we find that the N-emitters cover a wide range of stellar masses ($7.0 \lesssim \log(M_\star/M_\odot) \lesssim 11.8$). The highest-mass values should be treated with caution, as they correspond to LRDs, whose nature remains unclear. Excluding these objects, our sample spans a stellar mass range of $7.0 \lesssim \log(M_\star/M_\odot) \lesssim 9.8$.

\begin{figure*}[tb]
\includegraphics[width=0.5\textwidth]{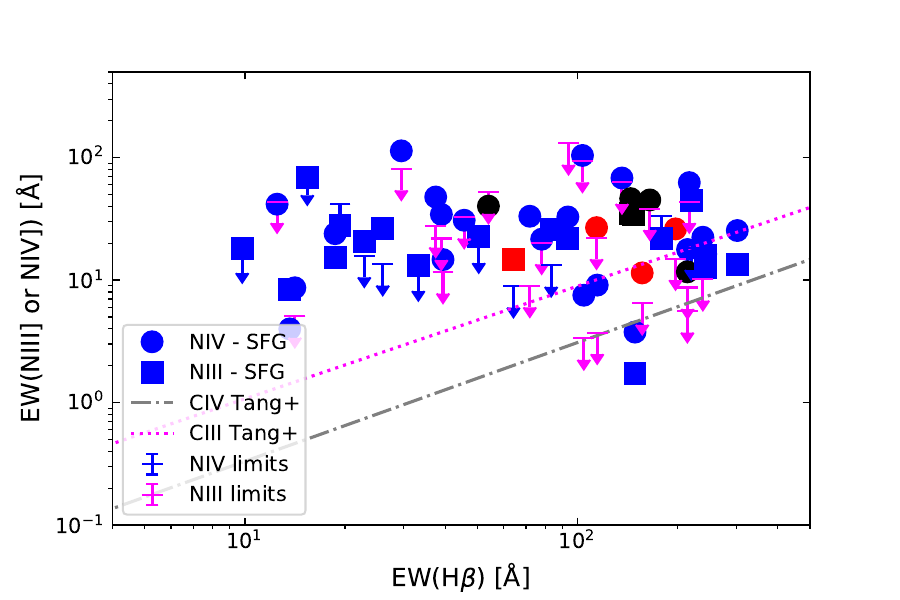}\vspace{-1mm}
\includegraphics[width=0.5\textwidth]{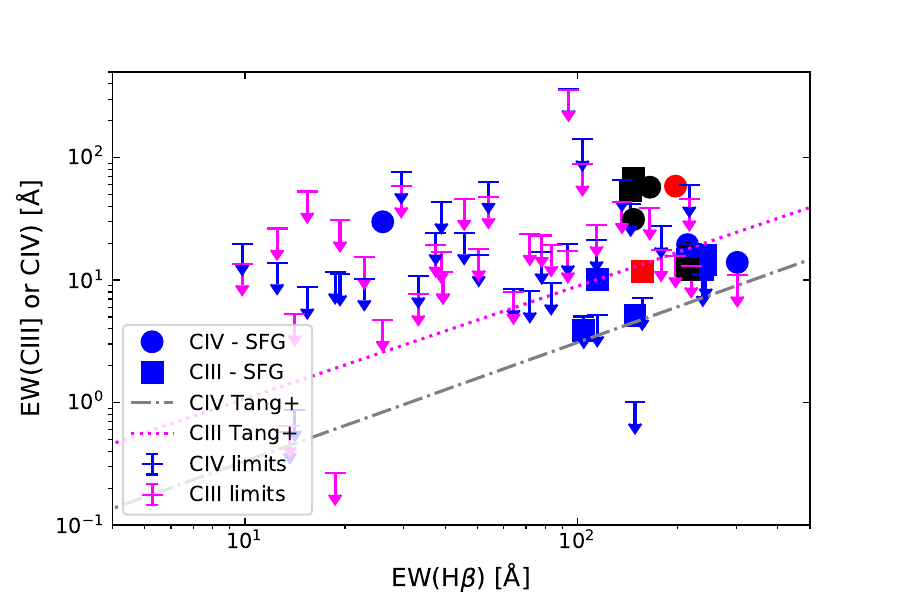}
\includegraphics[width=0.5\textwidth]{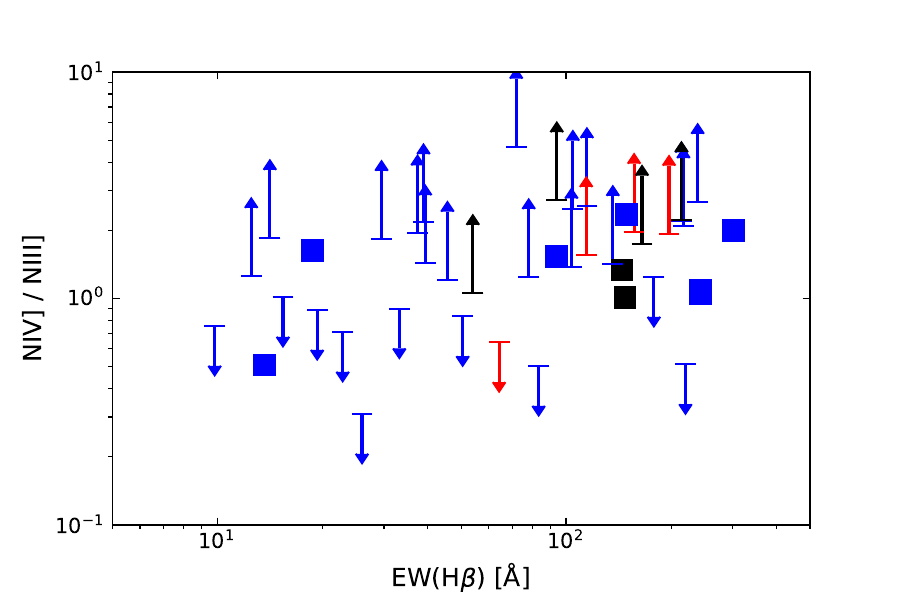}
\includegraphics[width=0.5\textwidth]{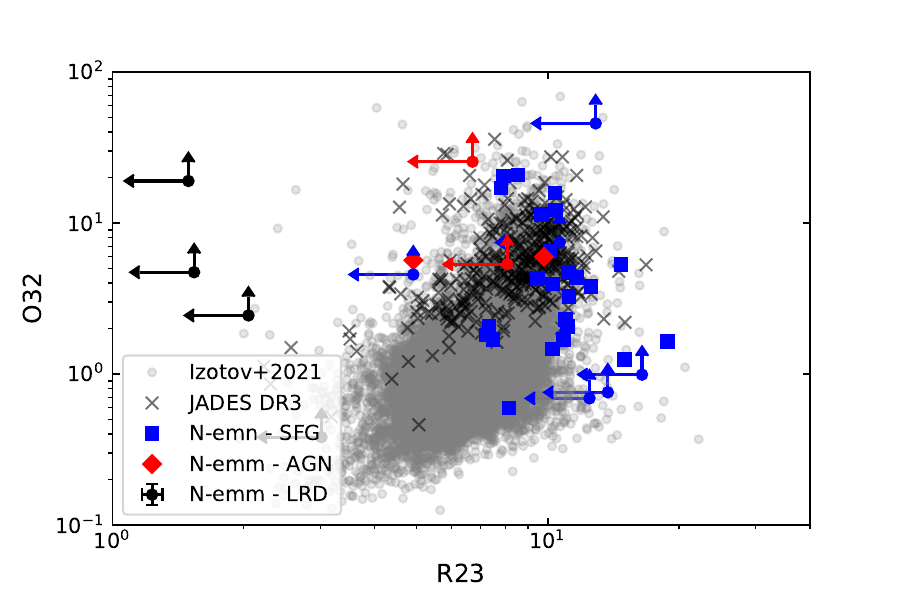}
\caption{Observed properties of N-emitters (SFG in blue, LRD in black, AGN in red). Equivalent widths of the UV lines (left: \Niiiuv, \Nivuv, right: \Ciiiuv, \Civuv) are shown as a function of EW(\hb) in the top row. 
Typical errorbars are comparable to the size of the symbols, and therefore not shown.
Upper limits for \Nivuv\ and \Civuv\ are shown as blue arrows, limits for  \Niiiuv\ and \Ciiiuv\ as magenta arrows.
Observed correlations describing Carbon-emitters, from \cite{Tang2025The-JWST-Spectr} are shown as dotted and dashed lines (for \Ciii\ and \Civ). Line ratios of \Nivuv/\Niiiuv\ versus EW(\hb), and O32 versus R23 are shown in the bottom panels. Low-$z$ and high-$z$ data from \cite{Izotov2021Low-redshift-co} and \cite{DEugenio2025JADES-Data-Rele} are shown in the bottom right panel. 
In all panels, $3 \sigma$ limits are shown. EWs are in the rest frame.}
\label{fig_uvlines}
\end{figure*}

\subsection{UV and optical emission lines of N-emitters}
\label{emission_lines}

The broad wavelength coverage provided by the PRISM spectra enables the detection of multiple emission lines.
In Table \ref{tab_lines} we report the rest UV and optical emission lines detected in our sample and their occurrence, i.e.~the number of detections in total, and the same information for the three N-emitter sub-categories (SFG, LRD, AGN).
The rest UV lines detected are \Nivuv, \Civuv, \Heiiuv, \Oiiiuv, \Niiiuv, and \Ciiiuv. 
Although not included in our measurements, \lya\ is detected in at least 12 sources. 

By selection, all objects show \Niiiuv\ or \Nivuv, and in 10 objects we significantly detect both of these lines (see Fig.~\ref{fig_uvlines}).
A larger fraction (74 \%) of N-emitters show \Nivuv, whereas $\sim 48$\% show \Niiiuv. \Ciii\ or \Civ\ lines are also detected, but only in a subset of our N-emitters. Finally, \Heiiuv\ and/or \Oiiiuv\ are detected in $\sim 24$ \% of the objects, and deblending of the lines is difficult. 
In the rest optical domain, the majority of objects show forbidden \Oii\ and \Oiii\ emission as well as H recombination lines, as expected (see Table \ref{tab_lines}). We detect, e.g., jointly \Oiii\ and one or several H recombination lines in 40 objects.
The remaining objects include two galaxies at $z>10$, where \Oiii\ is outside the spectral range covered by the observations, objects with Oxygen line detections but no H lines,  
and sources with low S/N in the optical.
Finally, we note that the auroral line \Oiiit\ is detected in two sources at high-$z$,
GN-z11 and in a $z=7.256$ source, which we classify as LRD (Table~\ref{tab_nemm}).


The strongest UV lines of our objects are generally \Niv\ or \Niii, although this is not a selection criterium. The \Civ\ and \Ciii\ lines are generally weaker or absent, but their equivalent widths (EWs) can reach comparable values of the N lines. 
Typical EWs of \Niv\ and \Niii\ are $\sim 10-50$ \AA. Similarly, for \Civ\ and \Ciii\ the EWs range from $\sim 4-30$ \AA\ and are somewhat higher in objects classified as AGN or LRD  (Fig.~\ref{fig_uvlines}). In some objects, we detect nitrogen lines with lower EWs $\la 10$ \AA, but our measurements are limited by S/N and the strong wavelength dependence of the spectral resolution of the PRISM, which facilitates emission line detections with increasing redshift.
In Fig.~\ref{fig_uvlines} we also compare our EW measurements to the typical/average values of those of \Ciiiuv\ and \Civuv\ in SFGs, which are known to scale with the \hb\ equivalent width. We clearly see that, compared to typical SFGs at $z \sim 6-9$ and at lower redshift, the EW of the nitrogen lines can exceed those of \Ciiiuv\ by a factor up to $\sim 10$.  The same is also true for the \Civuv\ line.

Interestingly, the equivalent widths of the optical lines (e.g.~\hb\ and \Oiiib, shown in Figs.~\ref{fig_uvlines} and \ref{fig_o32_ewo3}) of the N-emitters vary by more than one order of magnitude, from the highest values typically observed (EW(5007)$\sim 2000-3000$ \AA) 
to EWs as low as $\sim 100$ \AA\ (or EW(OIII+Hb) $\ga 130$ \AA).
The mean EW(OIII+Hb) $\approx 1100$ \AA\ (for our $z>5$ sources) is comparable to that reported by \cite[][$\sim 700-1000$ \AA]{Roberts-Borsani2024Between-the-Ext} from stacks.
Our results strongly contrast with those of \cite{Topping2025Deep-Rest-UV-JW}, who exclusively found N-emitters with very high equivalent widths. See discussion in Sect.~\ref{s_discuss}.

The relative intensities of the UV Nitrogen lines vary significantly, as shown in Fig.~\ref{fig_uvlines}.
Of course the dynamic range of the N43=\Nivuv/\Niiiuv\ ratio is somewhat limited, since these lines are relatively weaker (have lower EWs) than, e.g., the optical \oiii\ and \oii\ lines. 
We did not find a strong correlation of N43, e.g.~with redshift (see Fig. \ref{fig_n4n3}), or with the optical line ratio O32=\Oiiib/\Oii, but N43 possibly increases on average for objects with stronger lines (higher EW(\hb) -- see Fig.~\ref{fig_uvlines}).

The ionisation of Carbon, as traced by the ratio C43=\Civuv/\Ciiiuv,  behaves similarly to N43, although these lines are detected in less objects. For comparison, in stacked UV spectra of large samples \Ciiiuv\ is always found to be the strongest emission line (after \lya) -- hence C43 $<1$ -- and N lines very weak or absent \citep[cf.][]{Roberts-Borsani2024Between-the-Ext,Hayes2025On-the-Average-,Tang2025The-JWST-Spectr} at high-$z$, and at lower redshift ($z \sim 2-5$), as well known from ground-based observations \citep{Shapley03,Le-Fevre2015The-VIMOS-Ultra}.

The O32 ratio of our objects varies from $\sim 1$ to 30 and correlates with the observed rest-optical emission line strengths (e.g.\Oiiib\ or \hb), as illustrated in Fig.~\ref{fig_o32_ewo3}. The observed trend between O32 and EW(\oiii) of the N-emitters follows well that of SFGs at $z \sim 1-2$ and up to $z \sim 9$ from \cite{Tang2019MMT-MMIRS-spect,Tang2025The-JWST-Spectr}. It could be somewhat steeper than the relation found by \cite{Izotov2021Low-redshift-co} from SDSS galaxies at $z\sim 0-1$, also shown in this figure. The LRDs, where \Oii\ is consistently undetected, stand out as extreme outliers in this figure.

The classical O32 versus R23 diagram (R23$=$\Oii+\Oiii)$/\hb$), which to first order traces ionisation and metallicity, is shown in Fig.~\ref{fig_uvlines} for the N-emitters, and for other samples of star-forming galaxies, from the JADES DR3 \citep{DEugenio2025JADES-Data-Rele} and from \cite{Izotov2021Low-redshift-co}. The N-emitters span a similar range as the objects from the JADES catalogue. Interestingly, we find N-emitters over a wide range of R23, including also up to the maximum observed values, which suggests that these objects are not only found at very low metallicity, as previously reported.

\section{Physical properties of N-emitters and candidates}
\label{s_props}

\subsection{Metallicities (O/H abundances) of N-emitters}
\label{s_oh}
We first determine metallicities from the emission lines. Since auroral lines, in particular \Oiiit\ is only detected in two objects, we use the so-called strong line methods, for which we revert to the recent calibrations of \cite{Sanders2025The-AURORA-Surv} from the AURORA survey and a large compilation of JWST spectra. We primarily use the lines of \Oii, \Oiii\ and H lines (\ha, \hb, or \hg, depending on redshift), which can be used with the O2, O3, and $\hat{R}$ calibrations. We also use the new RO2Ne3 calibration, which involves the \Neiii\ line, and where RO2Ne3=(NeIII+\Oii)/$H\delta$. We replace the $H\delta$ flux by \hb\ or other stronger H lines, which can be measured more accurately and whose relative flux we set to the usual case B values. 
Among these calibrations, two are monotic, i.e.~single-valued, over the entire metallicity range covered by \cite{Sanders2025The-AURORA-Surv}: O2, RO2Ne3. To break the degeneracies of the remaining calibrations (O3, $\hat{R}$) we therefore combine them with O2 (O2=\Oii/\hb). We call these methods O3+O2 and $\hat{R}$+O2.
Note also that we do not use the R23 calibration, since the high-$z$ observations indicate that it is not sufficiently sensitive to O/H over a fairly broad range \citep[see][]{Sanders2025The-AURORA-Surv}. 
We have thus been able to determine metallicities for 28 objects with the O2 calibration, 15 with RO2Ne3, and 26 with O3+O2 and $\hat{R}$+O2. Finally, we adopt as our best estimate the mean of the \oh\ values (up to 4). 
We do not use metallicities for objects classified as LRDs or AGN, since the adopted calibrations are not appropriate for such objects.

As already hinted at by the range of R23 ratios covered by our objects (Fig.~\ref{fig_uvlines}), we find metallicities between $\oh \sim 7.15$ ($\sim 3$ \% solar) up to $\sim 8.5$ or even close to solar, depending on the indicator used.
A comparison of the metallicities derived with these calibrations and methods is shown in Fig.~\ref{fig_oh} as a function of \oh\ derived from the O2 calibration. Overall, the different metallicity indicators agree, but with a large scatter. At the low- and high-metallicity end, the O2 and RO2Ne3 methods differ more strongly with differences of $\sim 0.3$ typically. 

For 7 additional objects, which have measurements only in O3 (=\Oiiib/\hb), i.e.~significant detections in both lines but not in \Oii, we have found that the upper limits of O2 are compatible with the O/H estimate from the O3 calibration. We therefore adopt the O3 metallicities, which are found between $\oh = 7.15-7.93$.
In total we have thus 35 objects with estimated metallicities $\oh = 7.15-8.5$.

Given that our metallicity estimate includes the O2 calibration (when \Oii\ is detected), we note that O/H could be biased toward lower metallicity, since 
\Oii, which has a fairly low critical density ($\sim (1.3-4.8)\times 10^3$ \cmc\ for the two doublet lines), could be affected (decreased) by collisions. 
However, the observed O32 ratios of most of our objects are not particularly high, as shown in 
Fig.~\ref{fig_uvlines}, indicating that this effect might be limited.
Comparing our metallicity estimate for the three galaxies in common with the literature (GN-z11, GNz9p4, CEERS1019) we find that they agree within the quoted uncertainties, except for GN-z11, where we obtain $\oh=7.54$, which is $\sim 0.3$ dex lower than previous literature values \cite{Senchyna2024GN-z11-in-Conte,Ji2025Connecting-JWST}. From this, and from the differences illustrated in Fig.~\ref{fig_oh} we estimate that our typical uncertainty on $\log($O/H$)$ is $\sim 0.2$ dex. Higher quality (S/N) and medium resolution spectra, allowing also for the detection of auroral lines, are needed for more accurate abundance determinations.


\begin{figure*}[tb]
\includegraphics[width=0.5\textwidth]{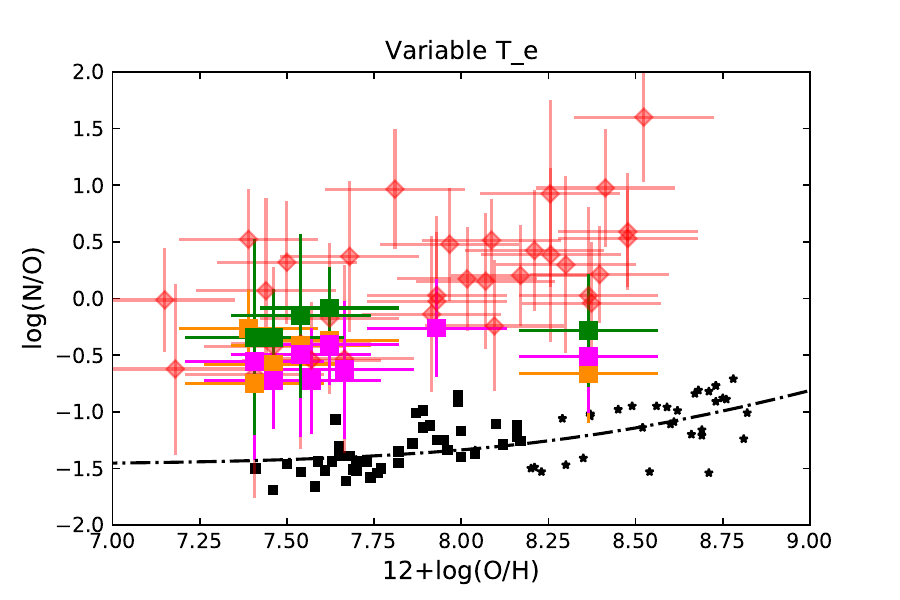}  \vspace{-1mm}
\includegraphics[width=0.5\textwidth]{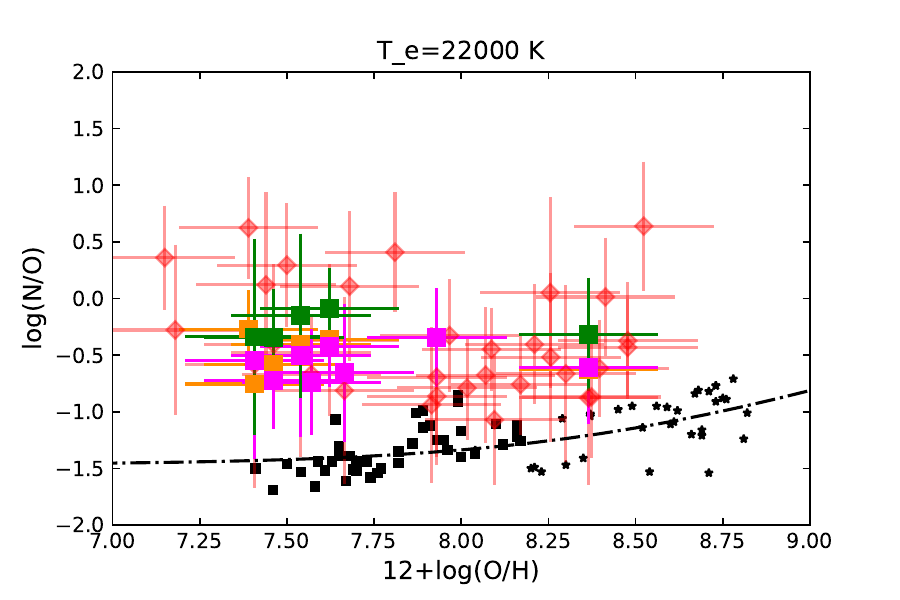}
\includegraphics[width=0.5\textwidth]{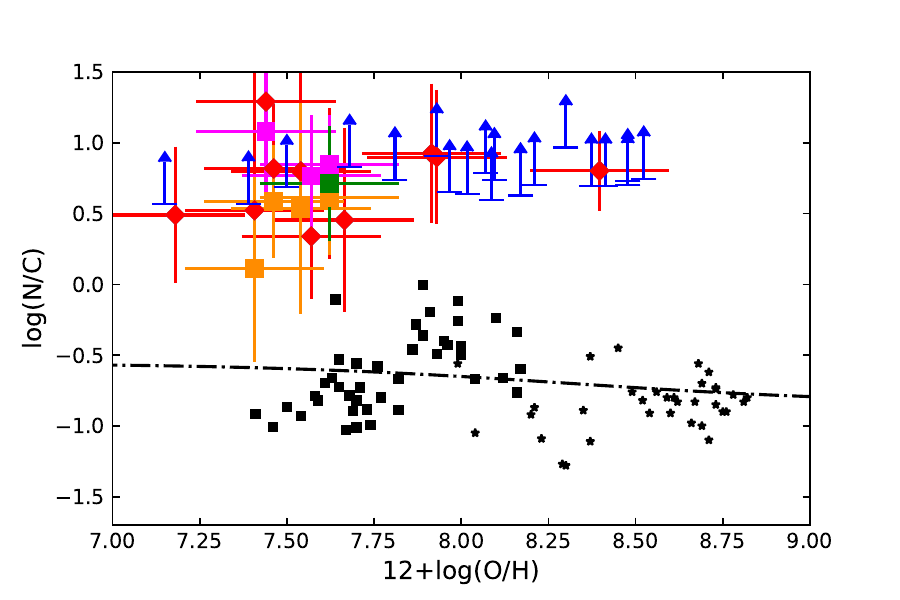}
\includegraphics[width=0.5\textwidth]{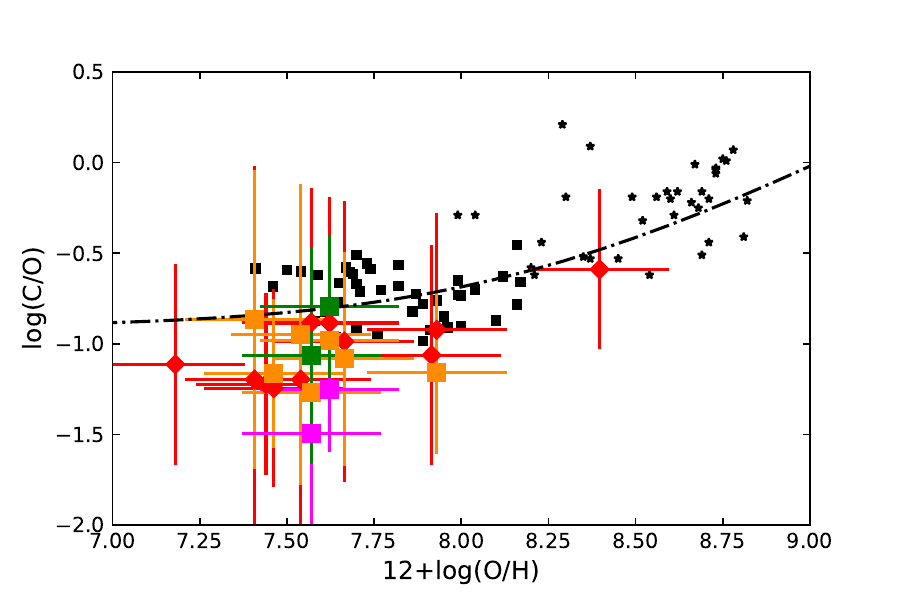}

\caption{Derived CNO abundance ratios of the N-emitters (colored symbols) as function of metallicity (O/H).
{\em Top left:} N/O abundances computed for variable $T_e(O/H)$, {\em top right:} for fixed $T_e=22000$ K.
Colored squares show N/O(UV) values from UV lines only (orange, magenta, and green for Eqs.~\ref{eq_nouv1}, \ref{eq_nouv1}, \ref{eq_nouv3}) and red diamonds show N/O(UV-opt) derived from UV and optical lines following Eq.~\ref{eq_opt}. 
{\em Bottom left:} N/C and lower limits from UV lines (blue), {\em bottom right:} C/O, using same colored symbols.
Low-$z$ star-forming galaxies and \hii\ regions from the compilation of \cite{Izotov2023Abundances-of-C} are shown by small black symbols, respectively. Dash-dotted lines show the average trend observed in low-$z$ star-forming galaxies, as parametrised by \cite{Vila-Costas1993The-nitrogen-to}  and \cite{Nicholls2017Abundance-scali} respectively.
}
\label{fig_no}
\end{figure*}

\subsection{CNO abundances of N-emitters}
\label{s_cno}

To analyse the CNO abundances of our sample we adopt several methods to determine the abundances of these elements. First, we only use UV emission lines to determine (when possible) relative abundances between C, N, and O, as this largely avoids corrections for attenuation and provides results which depend only very little on the (unknown/assumed) electron temperature and density.

Concretely, we determine the following ionic abundance ratios:
\begin{align}
        1.\ \frac{\rm N}{\rm O}(\rm UV) &\approx \frac{\rm N^{+2}}{\rm O^{+2}} \times {\rm ICF}=\frac{\rm N^{+2}}{\rm O^{+2}} \left[\frac{X(\rm N^{+2})}{X(\rm O^{+2})}\right]^{-1} \label{eq_nouv1} \\
        2.\ \frac{\rm N}{\rm O}(\rm UV)  &\approx \frac{\rm N^{+3}}{\rm O^{+2}} \times {\rm ICF}=\frac{\rm N^{+3}}{\rm O^{+2}} \left[\frac{X(\rm N^{+3})}{X(\rm O^{+2})}\right]^{-1} \\
        3.\ \frac{\rm N}{\rm O}(\rm UV)  &\approx \frac{\rm N^{+2} + N^{+3}}{\rm O^{+2}},\label{eq_nouv3}
\end{align}
(where $X$ stands for the ionisation fraction) from the \Niiiuv, \Nivuv, and \Oiiiuv\ lines, where the strength of \Oiiiuv\ is derived from the blend of \Heiiuv+\Oiiiuv, assuming a typical value of \Heiiuv/\Oiiiuv$=0.25$ following \cite{Tang2025The-JWST-Spectr}. This is applicable to 11 N-emitters from our sample for which the \Heiiuv+\Oiiiuv\ blend is significantly detected.
We then use the expressions of \cite{Villar-Martin2004Nebular-and-ste} to determine the ionic abundance ratios, and assume ionisation correction factors ICF$=1$.
According to the recent calculations of \cite{Martinez2025Under-Pressure:} the above ICF's are all $>1$ or provide corrections which are small ($\la 20$ \%) compared to our systematic uncertainties. Formally, our reported N/O values could therefore underestimate the true values.

We proceed in the same way to determine up to three estimates of the C/O(UV) abundance ratio, using the \Ciiiuv\ and \Civuv\ lines instead of the nitrogen lines. 
And we compute ionic abundance ratios of N/C (or lower limits thereof) using again only the relative line intensities of the UV lines (see Appendix~\ref{s_app_abundances}).
The main advantages of relative abundances of CNO elements derived from the UV lines is that these depend only weakly on the (unknown) electron temperature and are essentially independent of density (as long as $n_e \la 10^9$ \cmc), as described further in Appendix~\ref{s_app_abundances}. 

Finally, we also derive N/O from the combination of UV and optical lines, from which we obtain the ionic N$^{+2}$/H$^+$ and N$^{+3}$/H$^+$ abundances \citep[again following][]{Villar-Martin2004Nebular-and-ste}, and then N/O(UV-opt) from
\begin{equation}
 \frac{\rm N}{\rm O}(\rm UV-opt) \approx \frac{\rm N^{+2}/H^+ + N^{+3}/H^+}{\rm O/H},
\label{eq_opt}
\end{equation}
where O/H is the metallicity previously derived (Sect.~\ref{s_oh}). When a single N line is detected, we simply adopt the corresponding ionic abundance ratio (i.e.~ (N$^{+i}$/H$^+$)/(O/H), where $i=2$ or $3$ and ICF$=1$), without correction for unseen ionisation stages of N.
In the absence of measurements of $T_e$ for our objects, we adopt two cases to explore the importance of this effect. First, we adopt a variable $T_e=T_e($\oiii$)$, as derived empirically for galaxy samples with auroral line detections of \Oiiit, which show a well-known average relation between $T_e$ and metallicity (O/H), as shown, e.g.~in Fig.~\ref{fig_te_oh} for the sample of \cite{Izotov2021Low-redshift-co} from the SDSS and the JWST AURORA survey at $z \sim 2-3$ from \cite{Sanders2025The-AURORA-Surv}. We also adopt a minimum temperature of $T_e=15000$ K. Alternatively, we adopt a constant value of $T_e=22000$ K, which is a relatively high value, especially for galaxies with metallicities $\oh \ga 8.0$. 
This will lead to lower N/O values.


The relative N/O abundances of the N-emitters, derived for the two choices of $T_e$ and using the different methods explained above, are shown in Fig.~\ref{fig_no} as a function of the adopted metallicity. For comparison, we also show the measured abundances in \hii\ regions and star-forming galaxies at low-redshift and the average evolution of N/O-O/H.
Clearly, the N/O abundance ratio of all (or the majority) of our N-emitters is high compared to typical galaxies, and significantly super-solar\footnote{Solar ratios are $\log($N/O)$=-0.86$ and $\log($C/O)$=-0.26$, according to \cite{Asplund2009The-Chemical-Co}.}, independently of their metallicity and the adopted electron temperature.
For the objects where the abundances can be derived from the UV lines only, we find $\log($N/O(UV)) $\sim -0.75 - 0$, from the different methods, over a wide range of metallicities and quite independently of the assumed electron temperature. These N/O(UV) values are comparable to those found earlier for the known N-emitters \citep[see e.g.~compilation of][]{Ji2025Connecting-JWST}. 

Interestingly, several N-emitters reported in Fig.~\ref{fig_no} show indications for values $\log($N/O$)>0$  which could exceed those previously reported in an LRD (UNCOVER-45924) and an AGN (GS-3073 in the high-density region), with abundance ratios $\log$(N/O)$=0.20^{+0.06}_{=0.05}$ and $\log$(N/O)$=0.42^{+0.13}_{=0.10}$ respectively \citep{Ji2025Connecting-JWST}. Indeed, with the method combining UV and optical lines, we find several objects with N/O(UV-opt) exceeding the solar N/O value by factors $\ga 30$ (1.5 dex), i.e.~$\log$(N/O)$\ga 0.6$. The trend of increasing N/O with O/H obtained with this method (left panel of Fig.~\ref{fig_no}) is related to our coupling of the electron temperature with O/H (variable $T_e$). The comparison with the right panel shows the results obtained for a constant $T_e=22000$ K in the line-emitting region, namely relatively constant N/O values with O/H, and less extreme, but still super-solar values for the majority of objects. 


The relative N/C and C/O abundances of our N-emitters and the low-$z$ comparison samples are also shown in Fig.~\ref{fig_no}.
Quite independently of the method used, we find high N/C values (typically $\log($N/C)$\sim 0.3-1$), comparable to or somewhat higher than earlier findings \citep[cf.][]{2024MarquesChaves,Ji2025Connecting-JWST}. The lower limits, derived from UV line ratios when neither \Ciiiuv\ nor \Civuv\ are detected, also consistently show that nitrogen is strongly enhanced with respect to carbon in all N-emitters. 
The C/O ratios of the bulk of the objects are comparable to those in typical metal-poor star-forming galaxies, although there are hints for a somewhat lower C/O value in some objects. 
The C/O ratios found in our new N-emitters span a similar range to those found earlier \citep[see e.g.~][]{2024MarquesChaves,Ji2025Connecting-JWST}. 
It is difficult to derive accurate limits on C/O, i.e.~using the UV ratio of \Ciiiuv/\Oiiiuv, since few objects have \Oiiiuv\ (or \Heiiuv + \Oiiiuv) detections.

To summarise, we find that the majority, if not all N-emitters, show super-solar N/O abundances and a significant
enhancement of N/O and N/C compared to typical galaxies with the same metallicity (O/H). 
When two or more UV emission lines of C, N, or O are detected we can determine relatively accurate abundance ratios of these elements, for three reasons: 1) their emissivities depend in a similar fashion on the (unknown) electron temperature, 2) the critical density of emission lines are very high (e.g.~$>10^9$ \cmc\ for \Niiiuv\ and \Oiiiuv\ which yield the ionic ratio N$^{+2}/$O$^{+2}$, and $\sim 10^5-10^9$ \cmc\ for \Ciiiuv), and 3) the ionisation correction is quite small ($<0.2-0.3$ dex), in particular for N$^{+2}/$O$^{+2}$ and N$^{+2}/$C$^{+2}$  \citep{Martinez2025Under-Pressure:}.
From this, we found consistently high N/O and N/C ratios, and high lower limits for the latter, across a wide range of metallicities (O/H), from a few percent solar up to nearly solar. Since our metallicities are derived only using strong line methods, we consider these to be fairly uncertain. 
Better observations, allowing one to determine electron temperature and density, ideally from multiple tracers, will be needed to obtain more accurate metallicities.

\subsection{Morphologies of N-emitters}
\label{s_morpho}

We retrieved automatic morphological parameters from DJA, which were derived using SourceXtractor++ by fitting Sérsic models to JWST NIRCam images\footnote{ \url{https://dawn-cph.github.io/dja/blog/2024/08/16/morphological-data/}}. Effective radii are available for 28 of 45 N-emitters. They span a range from the unresolved detection limit ($R_{\rm eff}\lesssim 90\ \mathrm{pc} $) up to $R_{\rm eff}\approx2.7\ \mathrm{kpc}$. We also examined their stellar mass and SFR surface densities, derived from these morphological parameters for a subset of the N-emitters (see Appendix \ref{s_app_properties_SFR}) 
Visual inspection of the NIRCam images further reveals that these objects do not necessarily appear compact. We find a surprising diversity of morphologies among the N-emitters in our sample. In addition to compact sources, similar to those reported in previous studies such as GN-z11 \citep{2025Alvarez}, GHZ2 \citep{2024Calabro} or GN-z9p4 \citep{2024Schaerer}, we also identify clumpy objects, and possible spiral structures at $z\sim3.5$. 
(see Fig.\ref{fig_morpho}). 
For comparison, CEERS-1019 \citep{2024MarquesChaves} was also found to contain three compact, different clumps. Many of the objects in our sample exhibit extended shapes. We show four representative examples of these N-emitters in Fig. \ref{fig_morpho}. 

We do not observe any clear trend of morphological evolution with redshift. This result contrasts with earlier works \citep[]{2024Schaerer,2025Harikane,Topping2025Deep-Rest-UV-JW}, which reported high-redshift N-enhanced galaxies significantly more compact than typical star-forming galaxies, suggesting that both star formation and stellar mass are strongly concentrated in dense regions. The N-emission could come from one of the clumps of these objects.

\section{Discussion}
\label{s_discuss}

\subsection{Diversity among N-emitters}
\label{s_diversity}

In contrast to earlier findings, one of the salient results of our study is that N-emitters show a wide diversity in multiple properties. For example, N-emitters are found over a wide range of stellar masses and SFR (Sect.~\ref{s_sfr}), over a wide range of metallicities (Sect.~\ref{s_oh}), in very compact objects but also spatially-resolved galaxies with diverse morphologies (Sect.~\ref{s_morpho}), and in objects with very diverse optical emission line strengths (Sect.~\ref{emission_lines}).  
Previously, N-emitters identified with JWST were known to be limited to objects with the following properties: a relatively high UV brightness ($\muv \la -19.5$), low metallicity ($\oh \la 8.0$), compact (or unresolved) sizes, and very high equivalent widths in \Oiiib\ and/or \hb\ \citep[see e.g.][]{2024MarquesChaves,2024Schaerer,2025Harikane,Ji2025Connecting-JWST,Topping2025Deep-Rest-UV-JW}.
We now discuss how these findings can be reconciled and what these results tell us about the phenomenon and nature of N-emitters.

As shown in Sect.~\ref{emission_lines}, our sample of N-emitters spans a wide range of $\mathrm{EW(H\beta)}$ values. These peculiar objects are not only found at the highest EWs, as previously reported for known N-emitters, but also extend down to $\sim10$ \AA\ in the most extreme case, where $\mathrm{H\beta}$ is often very faint or barely detected. Interestingly, most of these low-EW sources appear spatially extended, except for two objects that are more compact but faint ($\muv \ga -17$) or display noisy spectra. 
Moreover, we identify clear Balmer breaks in several spectra, suggesting the presence of older stellar populations
(an example is shown in Fig.~\ref{fig_spec}).
We examined all galaxies showing a Balmer break and find that they also exhibit low $\mathrm{EW(H\beta)}$ values (up to 50\AA), indicating that N-emitters are not necessarily dominated by very young or recently formed stellar populations. 

\begin{figure}[tb]
\centering
\includegraphics[width=0.4\textwidth]{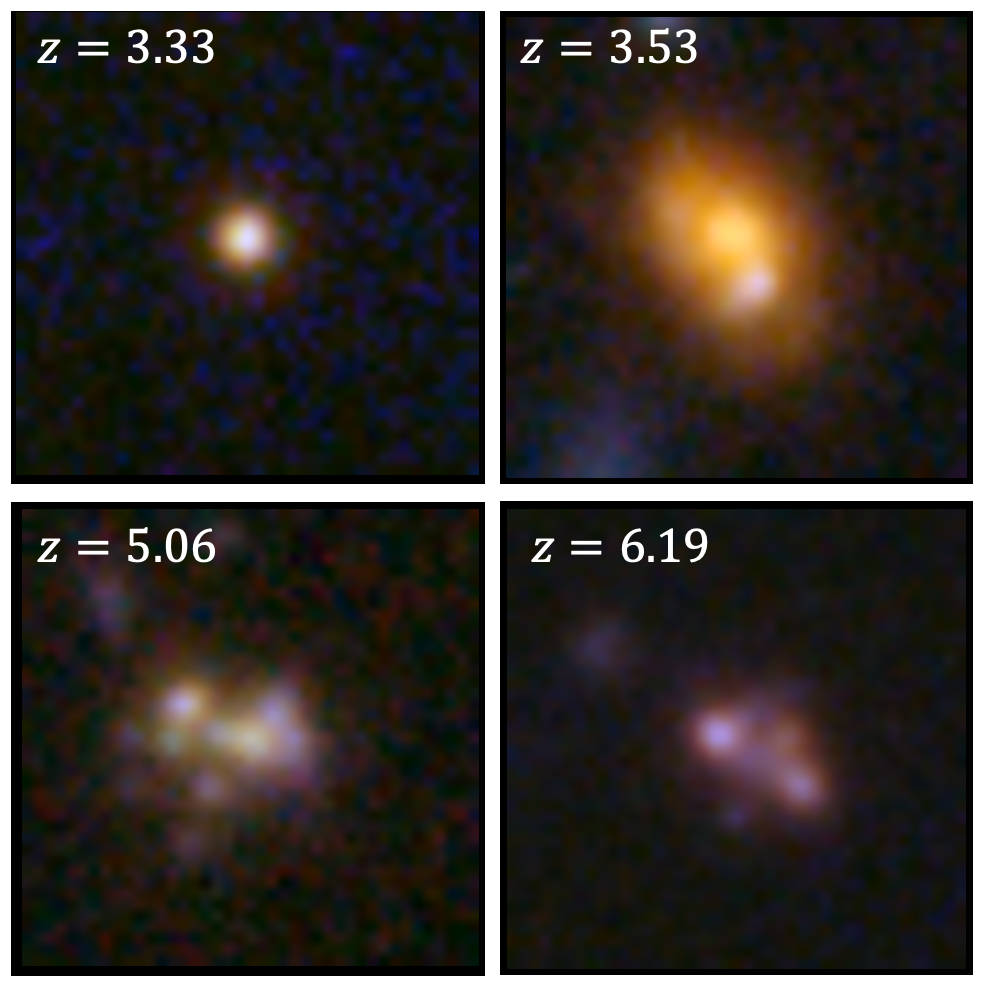}
\caption{NIRCam F115W + F356W + F444W composite $1\arcsec\times1\arcsec$ images of four N-emitters from our sample, illustrating the morphological diversity within the population, including compact, disky, and clumpy structures.}
\label{fig_morpho}
\end{figure}

The fact that the majority of low-EW(\hb) sources are extended and show colour variations suggests that they contain a mixture of stellar populations of different ages. In this case the (small, $\sim 0.2$ \arcsec) region where the JWST MSA spectrum was taken could include both a young star-forming region producing the bulk of the UV emission and the emission lines plus an older population ($\ga 50-100$ Myr), which is revealed by the Balmer break and ``dilutes'' (reduces) the EWs of the optical emission lines. The finding of Nitrogen lines in the UV with similar EWs, independently of EW(\hb) (Fig.~\ref{fig_uvlines}), also supports the presence of a young stellar population in all the objects. Therefore, we conclude that our observations indicate that strong N-emitting regions are likely relatively young ($\la 10$ Myr), although their exact age cannot be inferred from our data.

Similarly, the fact that many of our objects are spatially extended does not rule out earlier results suggesting that N-emitters were essentially found among those with very high stellar mass and SFR surface densities and very compact objects. The two observations can be reconciled if Nitrogen emission originates from a small localised region in the galaxy probed by the JWST aperture, which would show these ``extreme'' conditions. Future observations, e.g.~with integral field spectroscopy, can in principle verify this hypothesis by measuring the spatial extension of the N-emission and N/O enrichment, testing if the ISM densities in these regions are also as high as found in earlier N-emitters, and other properties. 

Finally, our finding of N-emitters over a wider range of metallicity and UV brightness than previous studies does not contradict them. We attribute this to our systematic search among a large number of spectra (i.e.~increased statistics; cf.~below), and the extension of our search towards lower redshifts, where both higher metallicity and UV-fainter galaxies can be found. 

\begin{figure}[tb]
\vspace*{-0.4cm}
\includegraphics[width=0.5\textwidth]{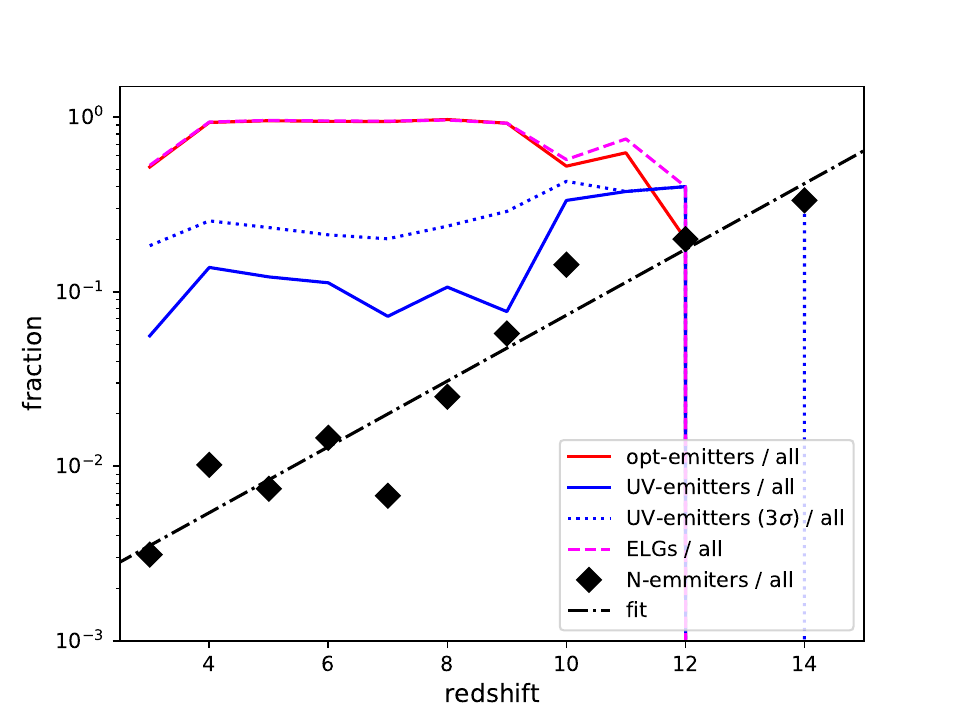}
\caption{Redshift evolution of the fraction of N-emitters (black diamonds), including both the literature and our sample, among all available NIRSpec PRISM spectra with robust redshifts (grade=3) from DJA. Blue lines show the fraction of rest-UV line emitters $4 \sigma$ (solid) or $3 \sigma$ detection (dotted) thresholds, respectively. The fraction of objects showing rest-optical emission lines is shown by the red line, and all emission lines galaxies (UV and/or optical) in magenta. At $z \protect\ga 11$ the number of objects is very small ($<10$ per bin) and few objects show emission lines.}
\label{fig_stats}
\end{figure}
\subsection{Statistical estimates of the N-emitter population}
\label{s_stats}

To compare the known and newly identified N-emitters with the total galaxy population and provide a first estimate of their relative fraction, we have used DJA to identify emission lines galaxies (ELGs) by searching for the presence of one or multiple emission lines with significance $\ge 4 \sigma$ in the rest-UV and the rest-optical domain, among all NIRSpec PRISM spectra with reliable redshifts (grade=3)\footnote{For UV lines we include \lya, \Heiiuv, \Oiiiuv, \Civuv, \Ciiiuv, \Nivuv\ and \Niiiuv; in the optical we search for \Oii, \Oiii, and H Balmer lines.}. We also count galaxies with no detectable emission line, and account for multiple entries of the same object in the database. The total sample includes 8323 unique sources above $z \ge 3$. The result is shown in Fig.~\ref{fig_stats}.

We find that the fraction of N-emitters, $f_N$, increases strongly with redshift,  when compared to all galaxies, galaxies with emission lines or objects showing just UV emission lines (``UV emitters'' in short). The trend is well described by 
\begin{equation}
\log(f_N)= 0.19 \times z -3.03.
\end{equation}
The fraction ranges from $\la 1$ \% of all galaxies with PRISM spectra at $z \la 6$ to more than 10 \% at $z\ga 10$. Since galaxies with emission lines dominate the total number of entries in DJA (at least over $3 \la z \la 10$), these values are also similar to those of N-emitters among emission line galaxies. Compared to galaxies showing UV emission lines, N-emitters represent $\sim 5-10$ \% at $z \sim 3-6$, depending on the adopted threshold (3 or 4 $\sigma$) for the automatic detection of UV lines. At $z \ga 10$ the fraction of N-emitters appears to be very elevated (up to $\sim 30$\%), although the statistics are low at these redshifts. 
At $z \ga 9$ the detection of rest-optical emission lines becomes more difficult, since the strongest lines (\Oiiib, in most cases) move out of the spectral range covered by NIRSpec. This explains the decrease of this population in Fig.~\ref{fig_stats}.
Since the DJA data used here come from a variety of programs with different selection criteria and depths, it is difficult to interpret precisely the fractions derived here. However, since both N-emitters and ``UV-emitters'' rely on the detection of UV lines and thus have similar detectability,
their relative fraction should be well determined and meaningful. And clearly, the fraction of N-emitters among ``UV emitters'' increases by a factor $\sim 6$ or more between $z \sim 3$ and 12.

The fact that N-emitters are rare, at least over redshifts $z \sim 3 - 10$ is not new, but it is quantified here for the first time. Indeed, large and deep ground-based spectroscopic surveys undertaken with the Keck and VLT have provided detailed rest-UV spectra for large numbers of galaxies from $z \sim 2-6$, showing no detection of \Nivuv\ or \Niiiuv\ in stacks (e.g. Shapley+, Du+2018, Lefevre+2019, Carilli+2021) and revealing very few peculiar objects with these features \citep[e.g.~GS-3073 at $z=5.55$, a N-emitter confirmed as AGN with JWST]{vanzella10,Ubler2023GA-NIFS:-A-mass}. Similarly, stacks of JWST spectra show no or weak emission \Niii\ or \Niv\ emission lines from $z \sim 4-9$, indicating that N-emitters are a rare phenomenon or population at these redshift \citep{Roberts-Borsani2024Between-the-Ext,Hayes2025On-the-Average-}. Finally, stacking the spectra of 10 galaxies at $z>10$, \cite{Roberts-Borsani2025JWST-Spectrosco} noticed the presence of \Nivuv\ in a subset of galaxies showing \Civuv\ emission, inferring also high N/O and N/C abundance ratios. These findings are fully consistent with our results of a strong redshift increase of the N-emitter fraction and a transition from rare to fairly common objects at $z>10$.

\subsection{On the nature of N-emitters}
\label{s_nature}
As mentioned in the Introduction, different scenarios have been proposed to explain the nitrogen enhancement observed in GN-z11 or the observed abundances in N-emitters discovered more recently. Since our work significantly expands the ``parameter space'' over which N-emitters are found and provides a first statistical estimate of the N-emitter population, we examine how this could constrain their nature or test these scenarios.

\subsubsection{Constraints from metallicity and relative N/C abundances}
So far N-emitters were only known at low-metallicity, with only one or two objects above $\oh \ga 8.0$ in the compilation of \cite{Ji2025Connecting-JWST}. If N-emitters also exist at higher metallicity, as our work suggests, metallicity-independent scenarios must be invoked. 
Furthermore, our finding of high N/C abundance ratios or lower limits ($\log($N/C$)\ga 0.5$ for the bulk of N-emitters) strongly constrains the scenarios. Our data indicate abundance ratios typical of ejecta from H-burning and no significant contribution of carbon, i.e.~no He-burning products, which are expected to be ejected in copious amounts by WR populations which include WC and/or WO stars.

For example, the models of \cite{Kobayashi2024Rapid-Chemical-} and \cite{Bekki2023A-model-for-GN-} for GN-z11 predict both an increase in N/O and C/O, leading to N/C values close to solar when the observed N/O value is reached. The N-emitter data is incompatible with this prediction.
In other models tracking the ejecta of normal stellar populations, a high N/O is reached after just $\sim 3$ Myr at low metallicities, and C/O increases shortly thereafter \citep[e.g.][]{Charbonnel2023N-enhancement-i,2024MarquesChaves,Shi2025Very-Massive-St}. To avoid high N/O and C/O and thus comply with our high N/C values, a narrow time-window ages ($\sim 3-4$ Myr only) or the absence of WC stars must be invoked \citep{2024MarquesChaves,2024Tapia,Shi2025Very-Massive-St}. For one individual case, e.g.~the Sunburst star cluster, this might be a valid explanation, but for a relatively large sample this is not plausible. 
Furthermore, since in normal stellar populations the ejecta of WR and WC stars 
increase with metallicity \citep[e.g.][]{Molla2012Modelling-the-c}, most N-enriched objects should also be C-enriched, which appears to be in contradiction with our finding of consistently high N/C and N/O ratios also at high metallicity. 

\subsubsection{Constraints from the statistics of N-emitters}
Our quantification of the N-emitter fraction and its evolution with redshift (Fig.~\ref{fig_stats}) could provide new hints on their nature and the sources of N-enhancement. We only briefly outline some ideas here and postpone a more detailed analysis to future work.

Naively, scenarios postulating that N-enhancement comes from the ejecta of WR stars would predict stronger enhancements and over longer periods with decreasing redshift, since the population of WR stars strongly increases with metallicity, which should increase on average over cosmic time. This would therefore lead to a redshift trend opposite to the observed one.
However, if N-emitters correspond to a brief and young phase, as suggested e.g.~by the GC scenario or pollution from massive stars (WR,VMS, or SMS) and by the finding of very high \hb\ equivalent widths \citep[][and our work]{Topping2025Deep-Rest-UV-JW}, the increase of $f_N$ with redshift could be explained by an increasing burstiness with $z$, which is supported by many observations and simulations \citep[e.g.][]{2025Cole}. If this effect is more important than the metallicity dependence discussed before, this could also reconcile the WR scenario with observations.
Alternatively, the formation of WR stars might also be facilitated at high-$z$, e.g.~due to rapid stellar rotation or IMFs favouring massive stars \citep[][]{Choi2017The-Evolution-a,Hutter2025ASTRAEUS:-X.-In}

On the other hand, if N comes mainly from older stellar populations, e.g.~AGB stars as proposed by \cite{DAntona2023,DAntona2025,McClymont2025The-THESAN-ZOOM}, N-enhancements would be expected in objects with lower EW(\hb), as shown in the simulations of \cite{McClymont2025The-THESAN-ZOOM}. 
If burstiness increases and the typical burst duration of galaxies decreases with redshift \cite[e.g.][]{Ciesla2024Identification-,2025Cole}, we would expect that such a scenario predicts a decreasing fraction $f_N$ towards high-$z$. Whether this holds up and if it can be used as a test of the AGB and differential outflow models remains to be examined in detail.

There are several ways to explain the observed increase of $f_N$ with redshift, for example:
{\em 1)} an IMF evolution, leading to a flatter IMF or a higher upper mass limit of the IMF, thus increasing the yields of massive stars \citep[WR or VMS, e.g.][]{Higgins2025The-impact-of-w}, and/or 
{\em 2)} an evolution of other conditions leading to an enhanced N production in compact high-density regions, such as the formation of supermassive stars in the core of very compact star clusters \citep[see][]{Gieles2018Concurrent-form,Gieles2025Globular-cluste}.
The latter, or probably a combination of the two, is a quite natural explanation, backed up by the following observations:
A shift towards higher SFR surface densities, $\Sigma_{\rm SFR}$, (on average) with increasing $z$ \citep[e.g.][]{2024Morishita}, the correlation between $\Sigma_{\rm SFR}$ and gas pressure in galaxy discs \citep[e.g.][]{Krumholz2012A-Universal-Loc}, and the increase of the average electron density with redshift \citep[e.g.][]{Isobe2023Redshift-Evolut,Martinez2025Under-Pressure:}, 
and the resulting/concommitant increase of the cluster formation efficiency (CFE), which favours the formation of massive gravitationally bound systems, such as proto-GCs \citep{Kruijssen2025The-Formation-o}. Last, but not least, this picture is also supported by the similarity between the observed abundance ratios in N-emitters and GCs \citep{Charbonnel2023N-enhancement-i,Senchyna2024GN-z11-in-Conte}, and the high formation redshifts of GCs.
If star-formation in these very dense clusters leads to WR, VMS, or even SMS, and allows or facilitates the retention of the enriched material within the cluster \citep[see simulations of][]{Shi2025Very-Massive-St}, this would naturally produce localised N-enriched regions with increasing frequency at high-$z$. This is our favoured scenario, which could potentially work with different nucleosynthetic sources (WR, VMS, or SMS), which remain to be identified.

\subsubsection{Other properties and constraints}
While the majority of N-emitters reported so far in the literature are very bright ($\muv \sim -22$ to $-19.5$, see Fig.~\ref{fig_muv}), the strongly lensed Sunburst cluster is considerably fainter ($\muv \approx -18.6$, adopting a magnification factor of $\sim 70$). Our sample confirms that N-emitters are found over a wide range of absolute scales, adding in particular also fainter objects (\muv\ down to $-17$ or fainter). Also, the host galaxies of our sample are found to span a wide range of stellar masses and SFR, showing thus that N-emitters are not limited to a specific mass or luminosity range.

As already discussed in Sect.~\ref{s_diversity}, our observations a priori relax the age constraints on N-emitters, since they show a wide range of \hb\ equivalent widths, in contrast to earlier, smaller samples which suggested N-emitters are associated only with very young ages \citep{Topping2025Deep-Rest-UV-JW}. However, our observations also show signs of composite stellar populations, which complicate age determinations and do not exclude that Nitrogen emission comes from a young region. Follow-up, spatially resolved observations will be key to answer this question and provide useful ages to constrain the source(s) of nitrogen.

Finally, we would like to comment on the fact that $\sim 13$\% (6 out of 45) N-emitters of our sample are LRDs, and several LRDs were already known to show UV emission lines of nitrogen \citep[e.g., UNCOVER-45924 and CANUCS-LRD-z8.6][]{Ji2025Connecting-JWST,Morishita2025A-Nitrogen-rich}. This, and the finding that other properties, such as compactness and the presence of unusually high-density gas, are found in (at least some) N-emitters and LRDs, shows that these objects share some relatively unusual physical properties. While the nature of the N lines and nucleosynthetic sources hosted by ``star-forming'' N-emitters and LRDs are unknown, one could speculate that the N-enrichment observed in both has the same origin. If this is true, and if LRDs host massive black holes, as generally postulated, it could suggest that some star-forming N-emitters contain supermassive stars, which can produce the observed ejecta and observed abundance ratios and subsequently collapse into intermediate-mass black holes \cite[see][]{Charbonnel2023N-enhancement-i,2024MarquesChaves}, thus forming the seeds for BHs in LRDs. Future studies exploring possible links between these two puzzling classes could shed more light on their nature.

\subsection{Caveats and open questions}

Finally, we briefly discuss some of the main caveats of our work, open questions, and possible future directions.

The physical interpretation of our results is subject to several uncertainties, primarily related to the derivation of chemical abundances and the limitations of the current dataset. 
First, the abundance estimates rely on dust-corrected fluxes and, as discussed in Section~\ref{s_extinction}, several sources in our sample exhibit negligible or even non-physical \ebv\ values possibly due to the limited spectral resolution of the PRISM data. This suggests that, on average, the derived \ebv\ values may be underestimated.
Nevertheless, this possible limitation does not affect our main conclusion. For higher extinction levels, the dust-corrected fluxes of the UV nitrogen lines (N\,{\sc iv}] and N\,{\sc iii}]) would increase more strongly than those of the optical oxygen lines used to compute N/O. As such, adopting higher extinction corrections would further increase the inferred N/O ratios. However, this does not affect the N/O ratios determined exclusively from UV lines.

A further source of uncertainty arises from the lack of direct $T_{\rm e}$ measurements. The [O\,{\sc iii}]~$\lambda$4363 auroral line is detected in only one source, preventing us from determining $T_{\rm e}$ for the bulk of the sample. As a result, O/H is derived using strong-line calibrations (Section~\ref{s_oh}), which are uncertain and could introduce systematic offsets. In addition, the electron densities of the N-emitting regions remain unconstrained due to the PRISM resolution, and thus, we cannot directly assess the possible presence of density gradients.
Nevertheless, the detection of lines with relatively low critical densities in the majority of sources (e.g., [O\,{\sc ii}]~$\lambda$3727, see Table~\ref{tab_lines}) suggests that strong collisional de-excitation is unlikely to dominate the observed line ratios. Therefore, strong line methods should provide a reasonable estimate of the metallicity (O/H) of N-emitters, at least on average. 
Furthermore, N/C and N/O ratios determined from UV lines are not very sensitive to the electron temperature and density, and should thus be relatively robust (see Appendix \ref{s_app_abundances}).
More accurate abundance determinations taking into account the likely presence of density and temperature gradients in N-emitters, as e.g.~shown by the detailed studies of \citep[][]{Martinez2025Under-Pressure:,Berg2025A-Fleeting-GLIM}, will require observations with higher S/N and spectral resolution.

To better characterise the physical properties of N-emitters, test different scenarios, and thus ultimately understand the nature of N-emitters, it will be key to spatially resolve these objects and map fundamental properties such as relative abundances, ISM properties (density and temperature), stellar populations, and ideally ionised gas kinematics. This is in principle feasible using, e.g., JWST NIRSpec IFU or with ELTs, for strongly lensed objects, 
and our study has identified further objects where spatially resolved studies are possible. 

\section{Conclusions}
\label{s_conclude}

We have carried out a systematic search for N-emitters, objects showing emission in one of the \Niiiuv\ or \Nivuv\ emission lines in the UV, using more than 20'000 JWST NIRSpec low-resolution (PRISM) spectra of $\sim 8300$ unique sources at $z >3$ available in the DAWN JWST Archive. Following an automatic preselection, tailored measurements, and careful inspection of the 2D and 1D spectra, we have identified 45 robust N-emitters between redshifts $z \sim 3-11$, including 4 previously known objects (GN-z11, CEERS-1019, GN-z9p4, and UNCOVER-45924). Our study effectively triples the number of known N-emitters.

Using the available spectra, multi-band images and photometric catalogs, we have then analysed the global properties of the N-emitters (morphology, stellar mass, star-formation rate etc.), characterised the observed spectroscopic properties (e.g., line strengths, excitation), estimated their metallicity (O/H abundance ratio), and determined the chemical abundance ratios of C, N, and O in these objects.
Finally, we have also determined, for the first time, a statistical measure of the population of N-emitters and its evolution with redshift. 

Our main results can be summarised as follows:
\begin{itemize}
\item N-emitters are found among a broad diversity of galaxies, in terms of morphology, UV magnitude, stellar mass, SFR, and metallicity. Morphologies range from very compact/unresolved objects, as previously identified, to extended galaxies showing multiple clumps, spatial colour variations and others. Stellar masses range from $\sim 10^7$ to $6 \times 10^9$ \msun, SFR $\sim 1-100$ \msunyr, and absolute UV magnitudes vary between $\muv \sim -16$ and $-22$ (Fig.~\ref{fig_muv}). 
\item Among the 45 N-emitters we classified six objects as Little Red Dots (LRDs), based on compactness, SED shape, and the presence of broad H lines, and four as broad-line AGN.
\item The UV nitrogen lines show typical equivalent widths between $\sim 5$ \AA\ and up to $\sim 100$ \AA\ in few cases. Diverse ionisation conditions, as traced by \Nivuv/\Niiiuv\ are observed. Carbon lines (\Civuv\ and \Ciiiuv) are generally fainter than the N lines (Figs.~\ref{fig_uvlines}).
\item The rest-optical emission lines of N-emitters show a wide range of ionisation and metallicities,  as traced by O32 and the relative intensities of \Oii, \Oiii, and \hb\ (R23, see Figs.~\ref{fig_uvlines}). Using different strong line calibrations established at high-redshift, we find metallicities $\oh \sim 7.15-8.5$, extending the metallicity range of $z>3$ N-emitters to metallicities above $\ga 20$ \% solar ($\oh \ga 8.0$) compared to earlier studies.

\item The \hb\ equivalent width of N-emitters also varies strongly, from EW(\hb)$\sim 300$ \AA\ to 10 \AA, and sources with low EWs show clear signs of a Balmer break, indicative of composite stellar populations combining both young ($\la 10$ Myr) stars responsible of the emission lines and an older population ($\ga 50-100$ Myr) contributing significantly to the rest-optical spectrum. 

\item The relative abundances of N, C, and O show clear signs of high N/O (supersolar values, $\log($N/O$) \sim -0.7$ to $0$ or higher) in all N-emitters. If the electron temperature in N-emitting regions decreases with increasing metallicity, the inferred N/O ratios could even increase with O/H. 
C/O abundances are ``normal'', comparable to those of galaxies at the same metallicity (O/H).
Interestingly, all N-emitters show high N/C ratios or lower limits ($\log($N/C$) \ga 0.5$), independently of metallicity. The observed abundance ratios are compatible with ejecta from H-burning (CNO equilibrium) and do not show signs of Carbon enhancements, even at higher metallicities (Fig.~\ref{fig_no}).

\item Comparing the N-emitter population with all emission lines galaxies we found a significant increase of the fraction of N-emitters with redshift, ranging from few times $10^{-3}$ at $z \approx 3$ to $10-30$ \% at very high-redshift ($z \ga 10$). We suggest that this is naturally explained, if N-enhancement traces overall young, compact star-clusters (possibly with IMFs favouring high-mass star), whose frequency increases with redshift (Sect.~\ref{s_nature}).
\end{itemize}

Overall, our study has increased the sample of known N-emitters by a factor $\sim 3$ and shown that their parameter space is larger than previously thought: the phenomenon of strong UV N emission is not only seen in very compact objects, not limited to very low metallicities, and not only found in objects with very strong emission lines (e.g.~high EW(\hb)), as suggested by earlier works \citep{2024MarquesChaves,2024Schaerer,2025Harikane,Topping2025Deep-Rest-UV-JW}. 
Despite this diversity, all N-emitters show enhanced N abundances (e.g. high N/O) -- which likely reflects the initial selection \citep[cf.][]{Zhu2025Only-Nitrogen-E} -- and interestingly also high N/C abundance ratios or lower limits. These results, and our finding of an increased N-emitter fraction with redshift, now provide new and broader observational constraints on the nature of N-emitters. We have confronted some of the existing scenarios with this data, showing, e.g., that the overall absence of C enhancement places stringent constraints on nucleosynthetic sources. 
It is the hope that this will allow future detailed studies, probably also including spatially-resolved information, to identify the main sources and physical processes governing the enigmatic N-emitters and understand their importance in the global picture of galaxy formation and evolution.

\bibliographystyle{aa}
\bibliography{references}

\begin{appendix}

\section{Emission line measurements and quality assessment}
\label{s_app_measure}
After examination of the 633 automatically-selected N-emitter candidates, we found that several of them are spurious (see Sect.~\ref{s_obs}).
We therefore remeasured the emission lines (fluxes, equivalent widths (EWs), and line widths (FWHM)) using two different approaches, which we describe here. 

First, we employed \texttt{Lime} 
\citep{Fernandez2024A&A...688A..69F_LIME}, a versatile tool that computes both integrated and Gaussian fluxes for a user-defined set of lines. As a preliminary step, we subtracted the rest-UV continuum of every source by modelling it with a power law of the form $f_{\lambda} \propto \lambda^{\beta_{\rm UV}}$, where $\beta_{\rm UV}$ is the UV slope. For this fit, we used the continuum spectral windows $\lambda_{\rm rest} = 1285$–$1385$\,\AA{ }and $\lambda_{\rm rest} = 2000$–$2300$\,\AA, which are relatively free of strong emission features. Given the low spectral resolution of PRISM, particularly in the rest-UV, and the proximity of the UV nitrogen lines to other emission features, we also simultaneously extracted fluxes for several other additional UV lines, including C {\sc iv} $\lambda$1550, He {\sc ii} $\lambda$1640, O {\sc iii}] $\lambda$1666, and C {\sc iii}] $\lambda$1908.

In the second approach, we developed a custom \texttt{Python} code to fit emission-line profiles using the \texttt{curvefit} package. As in the previous method, the fits were performed on continuum-subtracted spectra. We modelled the profiles of the nitrogen and other UV emission lines (as before) with single Gaussian components. The amplitudes and line widths were treated as free parameters, although for the latter we imposed a lower limit corresponding to the PRISM instrumental resolution. The centroid of each Gaussian was fixed to the expected observed wavelength, with the redshift of each source included as a free parameter and allowed to vary within $\pm 10\%$. At the same time, we extracted fluxes of several main rest-optical emission lines ([O\,{\sc ii}], [Ne\,{\sc iii}], H$\gamma$, H$\beta$, [O\,{\sc iii}], H$\alpha$). 
For rest-optical features, the underlying continuum was modelled (and subtracted) using spectral windows of $\Delta \lambda_{\rm rest} = 100$\,\AA{} on either side of each emission line, avoiding contamination from neighbouring emission lines.
For the uncertainties, we repeat the fitting process on 500 simulated spectra, introducing random noise to the observed spectrum. The noise is drawn from a Gaussian distribution with a standard deviation set by the 1$\sigma$ uncertainty spectrum. 

The measurements from these two methods were then used to construct the final N-emitter catalogue, as described in Sect.~\ref{s_obs}.

\begin{table}
    \centering
\caption{Significant line detections ($\ge 3 \sigma$) listing the number of objects and their separation into SFG, LRD, and AGN types.}
    \begin{tabular}{lcccc}
    Line & Detections & SFGs & LRDs & AGN \\
    \hline
        N IV] & 33 & 24 & 6  & 3 \\
        C IV & 8 & 4 & 3 & 1 \\
        He II & 3 & 3 & 0  & 0 \\
        O III] & 6 & 4 & 1 & 1 \\
        He II + O III] & 11 & 9 & 1 & 1 \\
        N III] & 21 & 18 & 2 & 1 \\
        C III] & 13 & 9 & 3 & 1 \\
        {[O II]} & 28 & 26 & 0 & 2 \\ 
        {[Ne III]} & 23 & 16 & 4 & 3 \\
        H$\delta$ & 13 & 8 & 2 & 3 \\
        H$\gamma$ & 26 & 18 & 5 & 3 \\
        {[O III]}$\lambda4363$ & 2 & 1 & 1 & 0 \\
        H$\beta$ & 34 & 24 & 6 & 4 \\
        {[O III]}$\lambda4959$ & 40 & 31 & 5 & 4\\
        {[O III]}$\lambda5007$ & 40 & 31 & 5 & 4\\
        H$\alpha$ & 38 & 29 & 5 & 4 \\ \hline 
        Total objects  & 45  &  35 &  6 & 4\\
        \hline
    \end{tabular}
\label{tab_lines}
\end{table}

\section{N-emitter sample}
\label{s_app_sample}

\subsection{Spectral classification: SFG, AGN, and LRD}
\label{s_app_agn}

To identify potential AGN, we first assessed whether the H$\beta$ and H$\alpha$ profiles show line widths significantly larger than the PRISM instrumental resolution, taking into account its wavelength dependence. 
Second, we compare the line widths of H$\beta$ to those of [O\,{\sc iii}]~$\lambda\lambda$4960,5008, since these lines are close in wavelength and thus have a similar instrumental resolution. Sources with H$\beta$ profiles significantly broader than [O\,{\sc iii}] lines are considered to have resolved H$\beta$ profiles, which, at the typical PRISM resolving power of $R \sim 100$, correspond to intrinsic FWHM values well exceeding $\sim 1000$\,km\,s$^{-1}$ (i.e., values that are commonly measured in type-I AGNs). According to this criterion, we identify ten sources exhibiting significantly resolved profiles in Balmer lines. Among these, six sources show strong Balmer breaks and V-like shapes in their PRISM spectra, features typical of Little Red Dots (LRDs, e.g., \citealt{Matthee2024ApJ...963..129M_LRDs}). We thus classify these six sources as LRDs. 

Finally, we also inspected the spectra for the presence of high-ionisation features (e.g., [Ne\,{\sc v}]~$\lambda$3346) and for UV and optical line ratios indicative of AGN activity. However, none of the sources exhibit these high-ionisation features, and the measured line ratios lack the depth required to robustly separate star-forming galaxies from AGN. We therefore conservatively classify all sources without broad Balmer lines ($N=36$) as star-forming galaxies (SFGs), while acknowledging that confirming the nature of their ionising sources will require deeper data and higher spectral resolution. A summary of the SFG, LRD, and AGN classifications for our sample is provided in Tables~\ref{tab_lines} and \ref{tab_nemm}.

\subsection{N-emitter sample}

Our complete sample of 45 N-emitters is listed in Table \ref{tab_nemm}. 

We also include one object serendipitously identified during the inspection of JWST PRISM spectra with grade=2 (uncertain redshift). This source (program ID: 6368 and slit ID: 43539) shows significant ($\gtrsim 3\sigma$) \Nivuv\ and \Civuv\ emission at $z \simeq 10.91$. Because no rest-optical lines fall within the PRISM wavelength coverage, its O/H and N/O abundances cannot be constrained.

\begin{table*}
    \centering
    \caption{N-emitter sample (45+1 objects), including 4 previously known N-emitters from the literature. 
    Note: $^{\dag}$ AGN. $^{\ddag}$ LRD. $^{*}$ Literature sample.}
    \begin{tabular}{l | c r r}
        DJA ID & Redshift & RA & DEC \\
        \hline
            capers-egs61-v4\_prism-clear\_6368\_43539.fits & 10.91 & 214.90311 &	52.80659 \\
jades-gdn2-v3\_prism-clear\_1181\_3991.spec.fits$^*$ & 10.61 & 189.10605 & 62.24205 \\
macsj0647-v3\_prism-clear\_1433\_3349.spec.fits & 10.17 & 101.97133 & 70.23972 \\
jades-gdn2-v2\_prism-clear\_1181\_3990.spec.fits$^*$ & 9.38 & 189.01700 & 62.24158 \\
rubies-egs52-nod-v3\_prism-clear\_4233\_8488.spec.fits$^*$ & 8.69 & 215.03539 & 52.89067 \\
capers-cos13-v4\_prism-clear\_6368\_183727.spec.fits & 8.29 & 150.08630 & 2.41956 \\
rubies-uds42-nod-v3\_prism-clear\_4233\_17998.spec.fits & 7.51 & 34.32391 & -5.29042 \\
j1007p2115-hennawi-v4\_prism-clear\_2073\_4562.spec.fits$^\ddag$  & 7.26 & 151.97823 & 21.28399 \\
jades-gdn2-v3\_prism-clear\_1181\_1893.spec.fits & 7.00 & 189.20531 & 62.25077 \\
rubies-egs61-v4\_prism-clear\_4233\_55604.spec.fits$^\ddag$  & 6.99 & 214.98303 & 52.95600 \\
capers-cos07-v4\_prism-clear\_6368\_35805.spec.fits$^\dag$  & 6.53 & 150.05536 & 2.29159 \\
rubies-uds42-v3\_prism-clear\_4233\_36171.spec.fits$^\ddag$  & 6.47 & 34.34500 & -5.26012 \\
capers-udsp2-v4\_prism-clear\_6368\_6267.spec.fits & 6.26 & 34.42320 & -5.11519 \\
jades-gds04-v3\_prism-clear\_1286\_166539.spec.fits & 6.19 & 53.09286 & -27.88261 \\
capers-egs53-v4\_prism-clear\_6368\_224070.spec.fits & 6.10 & 214.81476 & 52.84741 \\
mom-cos05-v4\_prism-clear\_5224\_318112.spec.fits & 6.06 & 150.09322 & 2.42429 \\
ceers-v2\_prism-clear\_1345\_403.spec.fits & 5.77 & 214.82896 & 52.87570 \\
gds-udeep-v4\_prism-clear\_3215\_212506.spec.fits & 5.35 & 53.15584 & -27.76672 \\
bullet-bradac-v4\_prism-clear\_4598\_7108588.spec.fits$^\ddag$  & 5.29 & 104.64603 & -55.96920 \\
mom-uds01-v3\_prism-clear\_5224\_162407.spec.fits & 5.25 & 34.39328 & -5.11823 \\
rubies-egs53-nod-v3\_prism-clear\_4233\_50052.spec.fits$^\dag$  & 5.24 & 214.82346 & 52.83028 \\
j0252m0503-hennawi-02-v4\_prism-clear\_2073\_2304.spec.fits & 5.21 & 43.05839 & -5.02353 \\
rubies-egs53-v4\_prism-clear\_4233\_44587.spec.fits & 5.11 & 214.82025 & 52.81237 \\
rubies-egs51-v3\_prism-clear\_4233\_16915.spec.fits & 5.06 & 215.07965 & 52.93826 \\
capers-cos13-v4\_prism-clear\_6368\_168143.spec.fits & 4.95 & 150.10789 & 2.42372 \\
mom-cos05-v4\_prism-clear\_5224\_312568.spec.fits$^\ddag$  & 4.92 & 150.10585 & 2.44682 \\
jades-gds-w05-v4\_prism-clear\_1212\_8712.spec.fits & 4.84 & 53.13946 & -27.84172 \\
jades-gds-wide3-v3\_prism-clear\_1180\_206035.spec.fits & 4.78 & 53.15817 & -27.78648 \\
mom-uds01-v3\_prism-clear\_5224\_165905.spec.fits$^\dag$  & 4.71 & 34.35557 & -5.11284 \\
gto-wide-uds10-v3\_prism-clear\_1215\_4257.spec.fits$^\dag$  & 4.63 & 34.57211 & -5.20520 \\
rubies-egs62-v3\_prism-clear\_4233\_63141.spec.fits & 4.60 & 214.82091 & 52.85705 \\
rubies-uds31-nod-v3\_prism-clear\_4233\_163999.spec.fits & 4.58 & 34.37717 & -5.11563 \\
gds-udeep-v3\_prism-clear\_3215\_200679.spec.fits & 4.56 & 53.11392 & -27.80620 \\
uncover-62-v3\_prism-clear\_2561\_44795.spec.fits & 4.47 & 3.60931 & -30.36787 \\
uncover-v4\_prism-clear\_2561\_45924.spec.fits$^*$$^\ddag$  & 4.46 & 3.58476 & -30.34363 \\
jades-gdn09-v4\_prism-clear\_1181\_54885.spec.fits & 4.45 & 189.23178 & 62.19730 \\
capers-cos07-v4\_prism-clear\_6368\_109397.spec.fits & 4.38 & 150.09329 & 2.28699 \\
jades-gdn11-v4\_prism-clear\_1181\_30362.spec.fits & 4.08 & 189.19444 & 62.27072 \\
capers-cos07-v4\_prism-clear\_6368\_33120.spec.fits & 3.77 & 150.08862 & 2.30459 \\
gto-wide-cos03-v4\_prism-clear\_1214\_3677.spec.fits & 3.76 & 150.14142 & 2.35504 \\
jades-gds04-v3\_prism-clear\_1286\_13938.spec.fits & 3.70 & 53.09673 & -27.88995 \\
rubies-uds41-v3\_prism-clear\_4233\_14460.spec.fits & 3.67 & 34.39204 & -5.29610 \\
cosmos-transients-v4\_prism-clear\_6585\_59719.spec.fits & 3.53 & 150.14154 & 2.35183 \\
jades-gds06-v3\_prism-clear\_1286\_283897.spec.fits & 3.47 & 53.08592 & -27.85871 \\
snh0pe-v4\_prism-clear\_4446\_511.spec.fits & 3.33 & 171.82788 & 42.48567 \\
rubies-uds42-v3\_prism-clear\_4233\_31658.spec.fits & 3.00 & 34.38974 & -5.26757 \\

        \end{tabular}
\label{tab_nemm}
\end{table*}

\begin{figure}[htb]
\includegraphics[width=0.5\textwidth]{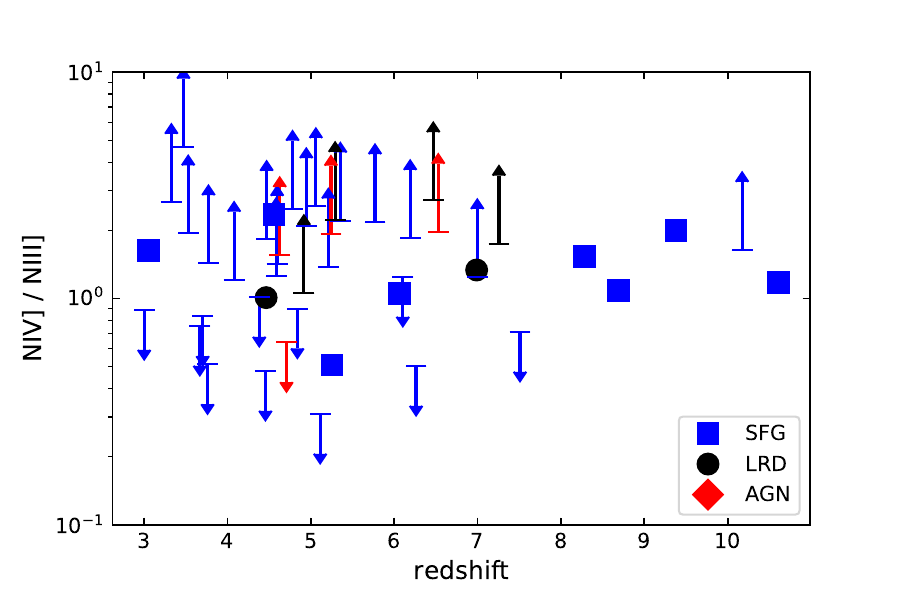}
\caption{Observed ratio of \Nivuv/\Niiiuv\ and 3-$\sigma$ upper and lower limits for the objects in our sample, as a function of redshift. SFGs, LRDs, and AGN are shown in blue, black, and red, respectively.}
\label{fig_n4n3}
\end{figure}

\begin{figure}[htb]
\includegraphics[width=0.5\textwidth]{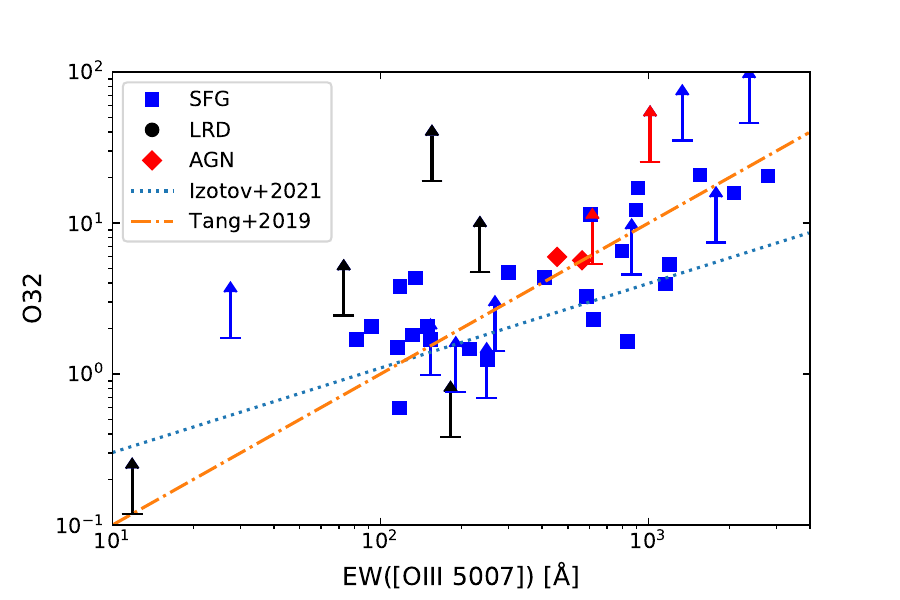}
\caption{Extinction-corrected ratio of \Oiiib/\Oii\ versus equivalent width of \Oiiib\ for our sample. Colours as in Fig.\~\protect\ref{fig_n4n3}. The dash-dotted and dotted lines show the average relations found for star-forming galaxies at $z \sim 1-2$ and low-$z$ by \cite{Tang2019MMT-MMIRS-spect} and \cite{Izotov2021Low-redshift-co}, respectively.
 }
\label{fig_o32_ewo3}
\end{figure}

\section{Observed and derived properties of N-emitters}
\label{s_app_properties}

\subsection{Emission line properties}
In Figs.~\ref{fig_n4n3} and \ref{fig_o32_ewo3} we show the observed ratio of \Nivuv/\Niiiuv\ as a function of redshift, and the O32 ratio as a function of the \Oiiib\ equivalent width. These figures complement those shown in Sect.~\ref{emission_lines}.
\begin{figure}[t]
\includegraphics[width=0.5\textwidth]{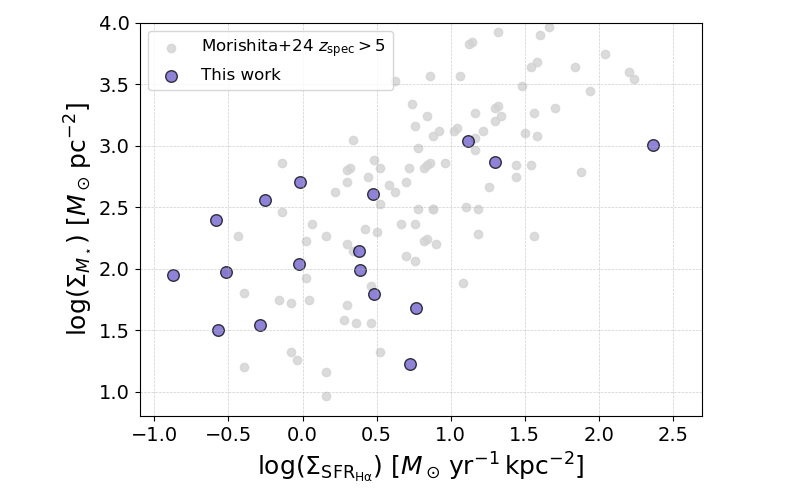}
\caption{Stellar mass and SFR surface densities for a subset of the new N-emitters for which size measurements and SED parameters are available. The grey points show a comparison sample of $z >5$ galaxies from \cite{2024Morishita}.}
\label{fig_SFRD}
\end{figure}

\subsection{SFR and stellar mass surface densities}
\label{s_app_properties_SFR}

Using the stellar masses, SFR, and galaxy sizes mentioned earlier (Sects.~\ref{s_sfr}, \ref{s_morpho}), we have computed the stellar mass and SFR surface densities of the N-emitters for which the data is available.
The resulting surface densities are shown in Fig.\ref{fig_SFRD} and compared to those of normal star-forming galaxies at $5<z<14$ derived by \citet{2024Morishita}. We find that N-emitters are not found exclusively at the highest values, in contrast to what was expected by, e.g., \citet{2024Schaerer} and \cite{2025Harikane}. 
Most of the objects lie below $\log(\Sigma_{M_{\star}}) = 3.0~\mathrm{M_{\odot}\,pc^{-2}}$
and $\log(SFR_{H_{\alpha}})=1.0~\mathrm{M_{\odot}\,yr^{-1}\,kpc^{-2}}$. Still, a few objects are found among the highest stellar mass and SFR surface densities, including CEERS-1019. This result supports the observed morphologies of the objects in our sample, as shown in Fig.\ref{fig_morpho}, indicating that N-emitters are not necessarily compact and are more diverse than previously found.

\section{Abundance determinations}
\label{s_app_abundances}

\subsection{Metallicity (O/H)}

\begin{figure}[tb]
\includegraphics[width=0.5\textwidth]{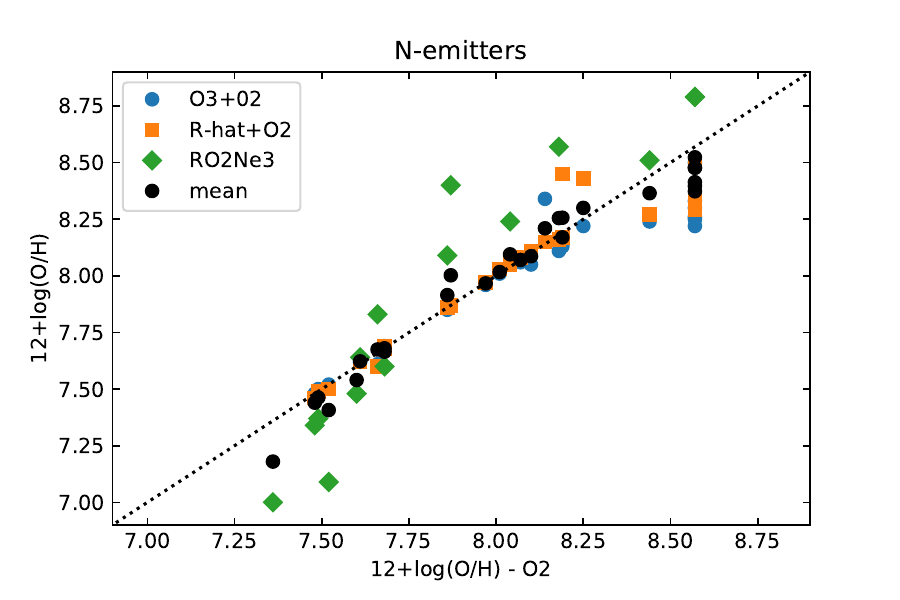}
\caption{Metallicity (O/H) comparison using different strong line calibrations and combinations (see text). The adopted metallicity is the mean obtained from different calibrations, shown by black circles.}
\label{fig_oh}
\end{figure}

Fig.~\ref{fig_oh} shows a comparison of the metallicities obtained with different strong line methods, and our final, adopted metallicity. This complements the text in Sect.~\ref{s_oh}.

\subsection{Comments on the determination of CNO abundances}

Illustrations providing further details on our abundance determinations are provided here.

\begin{figure}[tb]
\includegraphics[width=0.5\textwidth]{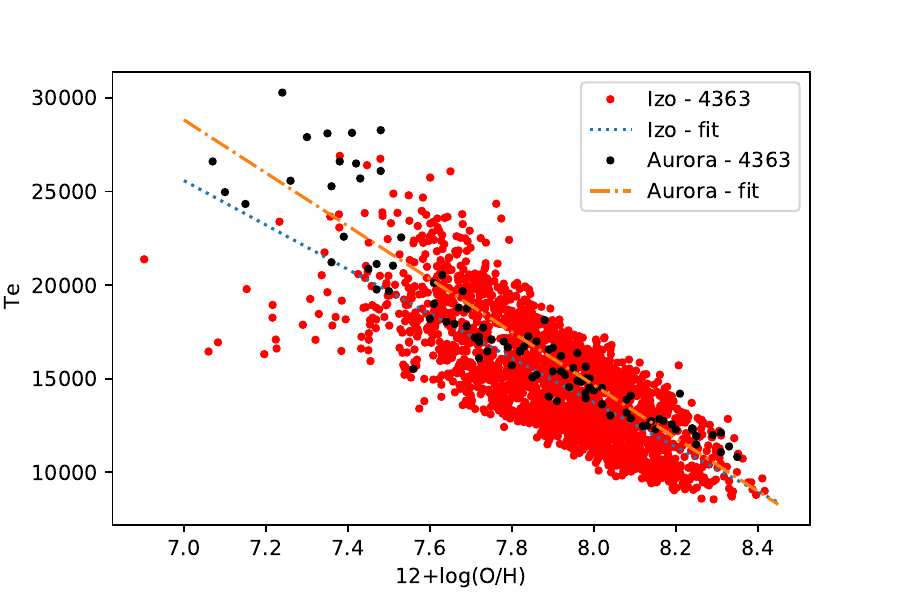}
\caption{$T_e($\oiii$)$ versus O/H obtained from the auroral line measurements in the low-redshift sample of \cite{Izotov2021Low-redshift-co} (red symbols), and JWST spectra of $z\sim 2-5$ galaxies from \cite{Sanders2025The-AURORA-Surv} (black points). The dotted and dash-dotted lines show a linear regression to the two samples.}
\label{fig_te_oh}
\end{figure}

In Fig.~\ref{fig_te_oh} we show the correlation between $T_e($\oiii$)$ and the O/H abundance, as derived from auroral line measurements of \Oiiit\ both in low- and high-$z$ star-forming galaxies. A linear fit to the AURORA data \citep{Sanders2025The-AURORA-Surv} yields $T_e(x)=-14172.25 *x + 128036.66 $ K, where $x=\oh$. We adopt $T_e($\oiii$)=max(15000,T_e(x))$ K for our abundance determinations with ``variable $T_e$'' in Sect.~\ref{s_cno}.

For relative N/C abundance, we similarly compute three versions,
\begin{align}
        1.\ \frac{\rm N}{\rm C}(\rm UV) &\approx \frac{\rm N^{+2}}{\rm C^{+2}} \times {\rm ICF} \\
        2.\ \frac{\rm N}{\rm C}(\rm UV)  &\approx \frac{\rm N^{+3}}{\rm C^{+3}} \times {\rm ICF} \\
        3.\ \frac{\rm N}{\rm C}(\rm UV)  &\approx \frac{\rm N^{+2} + N^{+3}}{\rm C^{+2} + C^{+3}},
\end{align}
where the last is in practice only applicable to one object where the four necessary lines are detected. 
For simplicity, we assume all $ICF=1$ here. According to the photoionisation models of \cite{Martinez2025Under-Pressure:}, the ICF for ${\rm N^{+2}}/{\rm C^{+2}}$ is between $0.8-0.95$ for ionisation parameters $\log(U) \sim -3$ to $-1$, which represents a small correction ($\la 0.1$ dex).

\subsubsection{On the temperature and density dependence of relative CNO abundances}

Key for ionic abundance ratio determinations are the line emissivities and their dependence on electron temperature and density, since by definition the abundance of an ion $X^i$ relative to hydrogen, e.g., is
\begin{equation}
    \frac{N(X^i)}{N(H^+)} = \frac{I_{\lambda(i)}}{I_{\rm H\beta}} 
                            \frac{j_{\rm H\beta}}{j_{\lambda(i)}},
\end{equation}
where $j$ are the emissivities, $I$ the line intensities, and we have chosen here the \hb\ line of hydrogen.
For the UV emission lines of interest here, the emissivities depend strongly on the electron temperature $T_e$, leading to considerable uncertainties when $T_e$ cannot be measured. 
For densities sufficiently below the critical density, one has typically
\begin{equation}
 \frac{N(X^i)}{N(H^+)} = c_i \sqrt{t}  E^0_{4,2} \times 10^{d_i/t} \times \frac{I_{\lambda(i)}}{I_{\rm H\beta}},
\label{eq_n3h}
\end{equation}
where $t=T_e/10^4$, and $E^0_{4,2}$ a term which weakly depends on $t$, and $c_i$ and $d_i$ are constants for each specific line $\lambda(i)$ \citep[see e.g.][]{Villar-Martin2004Nebular-and-ste}. 
For example, for N$^{+3}$/H$^+$ one has $c_i=1.06 \times 10^{-7}$ and $d_i=4.2$ for the relative line intensities of \Niiiuv\ and \hb.
As Eq.~\ref{eq_n3h} shows, the emissivity of \Niiiuv\ therefore increases rapidly with increasing $T_e$, leading to a lower ionic N$^{+3}$/H$^+$ abundance. To circumvent this strong temperature dependence, we have used the above relative ionic ratios between C, N, and O from UV lines, which all show strong, but comparable temperature dependencies (i.e.~large, but similar values of $d_i$).
For example, the ionic abundance ratio of 
\begin{equation}
    \frac{\rm N^{+2}}{\rm O^{+2}} \propto 10^{(d_i-d_j)/t} \times \frac{I_{\lambda(i)}}{I_{\lambda(j)}},
\end{equation}
derived from the relative intensities of \Niiiuv\ and \Oiiiuv\ is then only weakly dependent on $T_e$, since $d_i-d_j=-0.18$ \citep[cf.][]{Villar-Martin2004Nebular-and-ste}.
Also, for N$^{3+}$/O$^{2+}$ one has $d_i-d_j=0.45$, whereas the relative emissivities between the rest-UV lines of Nitrogen and \Oiiib\ depend very strongly on $T_e$. 
And ${\rm N^{+2}}/{\rm C^{+2}}$ and ${\rm N^{+3}}/{\rm C^{+3}}$ also show a weak temperature dependence.
In short, relative abundances of CNO elements derived from the UV lines depend only weakly on the (unknown) temperature, whereas abundance ratios involving hydrogen (and/or other combinations of UV and optical lines) are very sensitive to $T_e$, which may also vary across the line-emitting regions.

Regarding the density dependence, the situation is quite simple. All the UV lines of C, N, and O are doublets or multiplets of semi-forbidden lines where at least one of the components has a high critical density \citep[typically $n_{e,crit} \sim 10^9-10^{10}$ \cmc, see e.g.][]{Martinez2025Under-Pressure:}. In practice, as long as the electron density is $n_e \la 10^9$ \cmc\ the total emissivity of the \Nivuv, \Niiiuv, \Oiiiuv, \Ciiiuv\ doublets or multiplets (i.e.~the sum of the emissivities of all components) is essentially constant, i.e.~density-independent (and thus equal to their value at the low-density limit). The same holds for the emissivity of \Civuv. Since our PRISM spectra do not resolve the component of these doublets/multiplets, we can therefore easily apply the above equations, and our relative abundances do not depend on the electron density, as long as $n_e \la 10^9$ \cmc.

We have used \texttt{pyneb} \citep{Luridiana2015PyNeb:-a-new-to} and recent atomic data to verify the validity of the above statements. We have also compared the analytic expressions from \cite{Villar-Martin2004Nebular-and-ste} with the emissivities obtained with \texttt{pyneb}, finding overall good agreement, and deviations of less than $<20$\% in some cases, which validates our approach.

\end{appendix}

\begin{acknowledgements}
DS wishes to thank the IAP, Paris, and its staff for their hospitality during a stay where some of this work was done.
And many colleagues for interesting discussions on very massive, supermassive stars, and related topics.
We also thank Zorayda Martinez for sharing results from her work in electronic format.
\\
We thank the numerous teams of the observational programs used in this study, for developing these valuable data sets. The N-emitters identified in this study derive in particular from the following programs: 
1180, 1181, 3215 (PI: Eisenstein), 
1212, 1214, 1215, 1286 (PI: Luetzgendorf), 
1345 (PI: Finkelstein), 
1433 (PI: Coe), 
2073 (PI: Hennawi), 
2561 (PI: Labbe), 
4233 (PI: de Graaff), 
4446 (PI: Frye), 
4598 (PI: Bradac), 
5224 (PI: Oesch), 
6368 (PI: Dickinson), and 
6585 (PI: Coulter)
This work is based on observations made with the NASA/ESA/CSA James Webb Space Telescope. The data were obtained from the Mikulski Archive for Space Telescopes at the Space Telescope Science Institute, which is operated by the Association of Universities for Research in Astronomy,Inc., under NASA contract NAS5-03127 for JWST.
The data products presented herein were retrieved from the Dawn JWST Archive (DJA).DJA is an initiative of the Cosmic Dawn Center(DAWN), which is funded by the Danish National Research Foundation under grant DNRF140.

\end{acknowledgements}

\end{document}